\documentclass[pdflatex,sn-mathphys-num]{sn-jnl}
\usepackage{graphics}
\usepackage{graphicx} 
\usepackage{multirow} 
\usepackage{amsmath,amssymb,amsfonts} 
\usepackage{amsthm} 
\usepackage{mathrsfs} 
\usepackage[title]{appendix} 
\usepackage{xcolor} 
\usepackage{textcomp} 
\usepackage{manyfoot} 
\usepackage{booktabs} 
\usepackage{algorithm} 
\usepackage{adjustbox}
\usepackage{pdflscape}
\usepackage{algorithmicx} 
\usepackage{algpseudocode} 
\usepackage{listings} 
\usepackage{bold-extra}
\usepackage[normalem]{ulem}

\theoremstyle{thmstyleone}

\theoremstyle{thmstyletwo}

\theoremstyle{thmstylethree}

\begin{document}

\title[Article Title]{A large thermal energy reservoir in the nascent intracluster medium at a redshift of 4.3}

\author*[1]{\fnm{Dazhi} \sur{Zhou}}\email{dzhou.astro@gmail.com}
\author[2,1,3]{\fnm{Scott} \sur{Chapman}}
\author[4,5]{\fnm{Manuel} \sur{Aravena}}
\author[6]{\fnm{Pablo} \sur{Araya-Araya}}
\author[7,8]{\fnm{Melanie} \sur{Archipley}}
\author[9]{\fnm{Jared} \sur{Cathey}}
\author[10,11]{\fnm{Roger} \sur{Deane}}
\author[12,13]{\fnm{Luca} \sur{Di Mascolo}}
\author[14]{\fnm{Raphael} \sur{Gobat}}
\author[15,16]{\fnm{Thomas} \sur{Greve}}
\author[1]{\fnm{Ryley} \sur{Hill}}
\author[17]{\fnm{Seonwoo} \sur{Kim}}
\author[17,18,19]{\fnm{Kedar} \sur{Phadke}}
\author[1]{\fnm{Vismaya} \sur{Pillai}}
\author[20]{\fnm{Ana} \sur{Posses}}
\author[21]{\fnm{Christian} \sur{Reichardt}}
\author[4,5]{\fnm{Manuel} \sur{Solimano}}
\author[20]{\fnm{Justin} \sur{Spilker}}
\author[22]{\fnm{Nikolaus} \sur{Sulzenauer}}
\author[17]{\fnm{Veronica} \sur{Dike}}
\author[17,18,23]{\fnm{Joaquin} \sur{Vieira}}
\author[17]{\fnm{David} \sur{Vizgan}}
\author[1]{\fnm{George} \sur{Wang}}
\author[22]{\fnm{Axel} \sur{Wei\ss}}

\affil*[1]{\ubc}
\affil[2]{\orgdiv{Department of Physics and Atmospheric Science}, \orgname{Dalhousie University}, \country{Canada}}
\affil[3]{\orgdiv{National Research Council}, \orgname{Herzberg Astronomy and Astrophysics}, \country{Canada}}
\affil[4]{\orgdiv{Instituto de Estudios Astrof\'{\i}cos, Facultad de Ingenier\'{\i}a y Ciencias}, \orgname{Universidad Diego Portales, Av. Ej\'ercito 441, Santiago}, \country{Chile}}
\affil[5]{\orgdiv{Millenium Nucleus for Galaxies (MINGAL)}}
\affil[6]{\orgdiv{Departamento de Astronomia, Instituto de Astronomia}, \orgname{Geofísica e Ciências Atmosféricas, Universidade de São Paulo Rua do Matão 1226, Cidade Universitária, 05508-900, São Paulo, SP}, \country{Brazil}}
\affil[7]{Department of Astronomy and Astrophysics, University of Chicago, 5640 South Ellis Avenue, Chicago, IL, 60637, USA}
\affil[8]{Kavli Institute for Cosmological Physics, University of Chicago, 5640 South Ellis Avenue, Chicago, IL, 60637, USA}
\affil[9]{\ufastro}
\affil[10]{\orgdiv{Wits Centre for Astrophysics, School of Physics}, \orgname{University of the Witwatersrand, 1 Jan Smuts Avenue, 2000, Johannesburg}, \country{South Africa}}
\affil[11]{\orgdiv{Department of Physics}, \orgname{University of Pretoria, Hatfield, Pretoria 0028}, \country{South Africa}}
\affil[12]{\orgdiv{Kapteyn Astronomical Institute}, \orgname{University of Groningen, Landleven 12, 9747 AD, Groningen}, \country{The Netherlands}}
\affil[13]{\orgdiv{Université Côte d'Azur}, \orgname{Observatoire de la Côte d'Azur, CNRS, Laboratoire, Lagrange}, \country{France}}
\affil[14]{\orgdiv{Instituto de Física}, \orgname{Pontificia Universidad Católica de Valparaíso, Casilla 4059, Valparaíso}, \country{Chile}}
\affil[15]{Cosmic Dawn Center (DAWN), \country{Denmark}}
\affil[16]{DTU Space, Technical University of Denmark, Elektrovej 327, DK-2800 Kgs. Lyngby, Denmark}
\affil[17]{\illinoisastro}
\affil[18]{\illinoiscaps}
\affil[19]{NSF-Simons AI Institute for the Sky (SkAI), 172 E. Chestnut St., Chicago, IL 60611, USA}
\affil[20]{\tamu}
\affil[21]{School of Physics, University of Melbourne, Parkville, VIC 3010, Australia}
\affil[22]{\mpifr}
\affil[23]{\illinoisphysics}

\newcommand{\illinoisastro}{\text{Department of Astronomy, University of Illinois, 1002 West} \linebreak \text{Green St., Urbana, IL 61801, USA}}
\newcommand{\illinoiscaps}{\orgdiv{Center for AstroPhysical Surveys}, \orgname{National Center for Supercomputing Applications, 1205 West Clark Street, Urbana, IL 61801}, \country{USA}}
\newcommand{\illinoisphysics}{\orgdiv{Department of Physics}, \orgname{University of Illinois, 1110 West Green St., Urbana, IL 61801}, \country{USA}}
\newcommand{\ufastro}{\text{Department of Astronomy, University of Florida, Gainesville, FL 32611, USA}}
\newcommand{\tamu}{\text{Department of Physics and Astronomy and George P. and Cynthia} \linebreak \text{Woods Mitchell Institute for Fundamental Physics and Astronomy, }\linebreak \text{Texas A\&M University, 4242 TAMU, College Station, TX 77843-4242, USA}}
\newcommand{\ubc}{\orgdiv{Department of Physics and Astronomy}, \orgname{University of British Columbia, 6225 Agricultural Rd., Vancouver, V6T 1Z1}, \country{Canada}}
\newcommand{\mpifr}{\text{Max-Planck-Institut für Radioastronomie, Auf dem Hügel 69,}\linebreak \text{53121, Bonn, Germany}}

\abstract{
Most baryons in present-day galaxy clusters exist as hot gas ($\boldsymbol{\gtrsim10^7\,\rm}\mathrm{K}$), forming the intracluster medium (ICM) \cite{Bryan1998}. 
Cosmological simulations predict that the mass and temperature of the ICM rapidly decrease with increasing cosmological redshift, as intracluster gas in younger clusters is still accumulating and being heated \cite{chiang2020,li2023,Rohr2025}.  
The thermal Sunyaev-Zeldovich (tSZ) effect arises when cosmic microwave background (CMB) photons are scattered to higher energies through interactions with energetic electrons in hot ICM, leaving a localized decrement in the CMB at a long wavelength \cite{Sunyaev1980,Voit2005}. 
The depth of this decrement is a measure of the thermal energy and pressure of the gas \cite{Mroczkowski2019}. 
To date, the effect has been detected in only three systems at or above $z\sim2$, when the Universe was 4 billion years old, making the time and mechanism of ICM assembly uncertain \cite{Mantz2018,Gobat2019,mascolo2023}. 
Here, we report observations of this effect in the protocluster SPT2349$-$56 with Atacama Large Millimeter/submillimeter Array (ALMA). 
SPT2349$-$56 contains a large molecular gas reservoir, with at least 30 dusty star-forming galaxies (DSFGs) and three radio-loud active galactic nuclei (AGN) in a 100-kpc region at $z=4.3$, corresponding to 1.4 billion years after the Big Bang
\cite{Miller2018,Hill2020,Chapman2024,Zhou2025}. 
The observed tSZ signal implies a thermal energy of $\mathbf{\sim 10^{61}\,\mathrm{erg}}$, exceeding the possible energy of a virialized ICM by an order of magnitude. 
Contrary to current theoretical expectations \cite{Sunyaev1972,li2023,Rohr2025}, the strong tSZ decrement in SPT2349$-$56 demonstrates that substantial heating can occur and deposit a large amount of thermal energy within growing galaxy clusters, overheating the nascent ICM in unrelaxed structures, two billion years before the first mature clusters emerged at $\mathbf{z \sim 2}$.  
}

\maketitle

SPT2349$-$56 was selected as the brightest protocluster candidate from the 2,500 deg$^2$ South Pole Telescope survey \cite{Reuter2020,Wang2021}. 
It is an active protocluster with a massive dark matter halo ($\sim10^{13}\rm M_\odot$) and a star-formation rate (SFR) of $\sim5,000\rm M_\odot$/yr within the central 100\,kpc \cite{Miller2018,Hill2020}. 
At least three radio-loud active galactic nuclei (AGN) and 30 dusty star-forming galaxies (DSFGs) are spectroscopically confirmed to reside in its core region \cite[][Chapman et al.\,in prep]{Miller2018,Hill2020,Chapman2024}, which may provide substantial energy injection into its potential nascent intracluster medium (ICM) \cite{Tornatore2003,Kravtsov2000,Conroy2008,McCarthy2008,Henden2019,Kooistra2022,Bennett2024,Chapman2024,Dong2023,Dong2024,Gardner2024,Bigwood2025,flamingosz_2025}. 
A large molecular gas reservoir has been detected in SPT2349$-$56 through low-resolution observations by the Atacama Compact Array (ACA) \cite{Zhou2025}, which might represent the cooler component of a proto-ICM, making it an ideal target to search for the thermal Sunyaev-Zeldovich (tSZ) signal. 

We obtained deep Band-3 observations at $3\rm\,mm$ with the Atacama Large Millimeter/submillimeter Array (ALMA) 12-m and ACA 7-m arrays and combined them with archival data. 
After subtracting the dust continuum emission from DSFGs in the protocluster using a Fourier-space technique, a strong, extended decrement is present in the core region (Methods).
The tSZ decrement peaks at 8.4$\sigma$ in the image and 10.4$\sigma$ in Fourier space, with an integrated flux density of --157$\pm$16\,$\mu$Jy.
This signal directly traces inverse Compton scattering of cosmic microwave background (CMB) photons by hot ICM. 
We quantify the signal strength by the Compton-$y$ parameter, defined as the fractional energy change of line-of-sight CMB photons, which is proportional to the integrated gas pressure along the line of sight \cite{Sunyaev1972,Carlstrom2002,Mroczkowski2019}. 
The corresponding integrated value within the system is further defined as Compton-$Y$ \cite{Sunyaev1980}. 
For SPT2349$-$56, the detected tSZ decrement corresponds to a Compton-$y$ parameter of $(5.6\pm0.8)\times10^{-6}$, or a Compton-$Y$ parameter of $(2.0\pm0.2)\times10^{-6}\rm\,arcmin^2$.

Because the tSZ effect is directly linked to the number density and temperature of hot electrons in a system, the ICM's thermal energy can be estimated \cite{Spacek2016}. 
From our measured Compton-$Y$ value, we infer a total thermal energy $E_{\rm therm}\approx10^{61}\rm\,erg$ (Methods). 
Historically, assuming shock waves as the primary heating source for a $10^{13}M_\odot$ protocluster at $z=4$, \citet{Sunyaev1972} originally predicted a $\sim10^{6}\,\rm K$ ICM, equivalent to a specific thermal energy of $\sim10^{14}\rm\,erg/g$.
To reproduce the observed decrement in this scenario, a gas mass of $5\times 10^{13}M_\odot$ is required, five times the total mass of the system. 
Even assuming a fully collapsed system with an abundant ICM \cite[$M_{\rm ICM}/M_{\rm halo}\lesssim0.06$,][]{Chapman2024}, the expected decrement would be only $Y\lesssim 4\times10^{-7}\rm\,arcmin^2$, corresponding to a distortion signal of $\lesssim30\,\rm\mu Jy$ (Methods). 
The measured tSZ signal thus exceeds these theoretical expectations by at least a factor of five, suggesting a large thermal energy reservoir in the forming ICM of SPT2349$-$56. 

A useful way to characterize the tSZ excess independent of cosmological redshift evolution is to compare it against the universal relation between cluster mass $M$ and the Compton-$Y$ parameter \cite{Arnaud2010}. 
The ICM in galaxy clusters is known to evolve self-similarly over the past 10 billion years, resulting in a tight correlation between $M$ and $Y$ once the redshift dependence $E(z)^{2/3}$ is factored out \cite{Arnaud2010,Maughan2012,plancksz,McDonald2017,Henden2019,Mostoghiu2019}. 
Although this relation accurately describes most tSZ-detected systems \cite{Marrone2012,Nagarajan2019,Bocquet2019,mascolo2023,Gobat2019,Marrewijk2024,Andreon2023}, the Compton-$Y$ in SPT2349$-$56 exceeds it by a factor of five (Extended Data Fig.\,\ref{fig:summary}), implying a breakdown of the ICM self-similarity at an epoch prior to its full virialization. 

\begin{figure}[htp]
    \centering
    \includegraphics[width=0.98\linewidth]{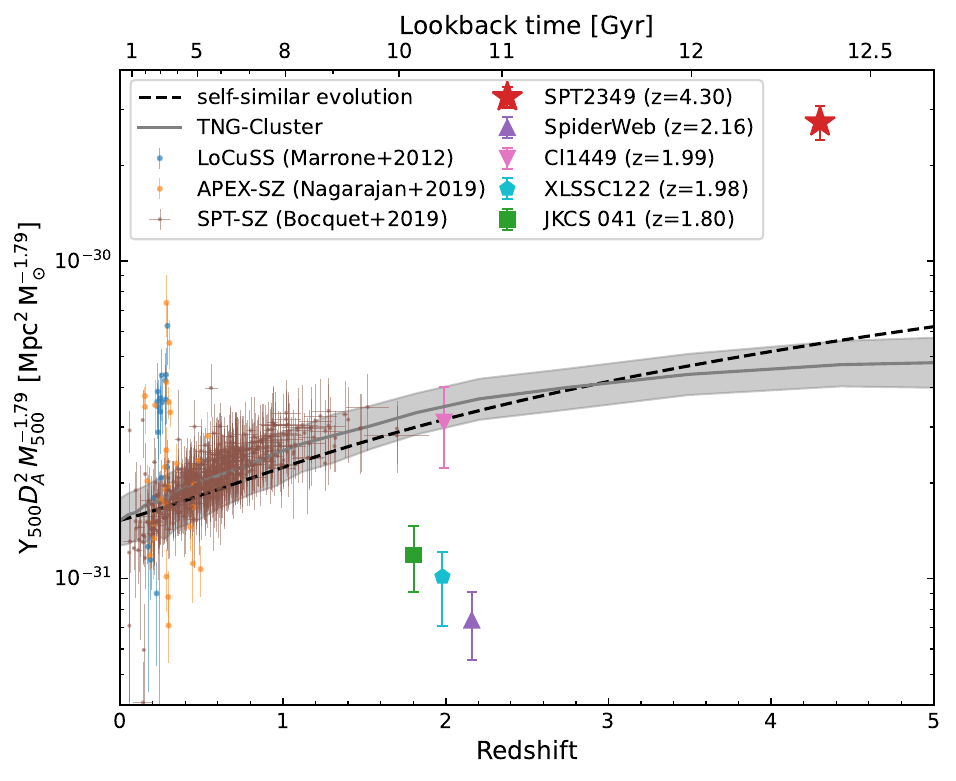}
    \caption{{\bf Redshift evolution of the mass-normalized Compton-$Y$ parameter. }
    The Compton-$Y_{500}$ is scaled by $\rm M_{500}^{1.79}$ to remove the mass dependence, following the universal tSZ scaling relation $Y\propto M^{1.79}$\cite{plancksz}. 
    The data points contain SPT-SZ \cite{Bocquet2019}, APEX-SZ \cite{Nagarajan2019}, and LoCuSS \cite{Marrone2012} results, along with high-$z$ systems detected in tSZ \cite{Mantz2014,Gobat2019,mascolo2023,Marrewijk2024}, including SPT2349$-$56 at $z=4.3$, shown as the red star. 
    The self-similar expectation is shown as the dashed line. 
    The solid gray line and shaded region represent the median and 68\% scatter from the TNG-Cluster simulation, which has been calibrated to the local $M_{500}-Y_{500}$ relation \cite{plancksz}. 
    The tSZ signal in SPT2349$-$56 lies 6.4$\sigma$ above the predicted evolutionary tracks, in contrast to more massive systems at later epochs.
    }
    \label{fig:z_evo}
\end{figure}

Indeed, because star formation and AGN activities were more intense and the ICM was less evolved at earlier epochs, one expects an increased scatter in the $M-Y$ relation at higher redshifts
\cite{Henden2019, li2023, Rohr2025, Bigwood2025, flamingosz_2025}. 
To further test this hypothesis, we used the TNG-Cluster simulations, a cosmological simulation suite of 352 zoom-in galaxy clusters, to predict the redshift evolution of the mass-normalized Compton-$Y$ (Methods). 
Fig.\,\ref{fig:z_evo} shows that the mass-normalized $Y$ from TNG-Cluster follows the self-similar prediction in the past 12 billion years, in line with current observations \cite{Marrone2012,Nagarajan2019,Bleem2015,Bocquet2019}, whereas tSZ decrement in less evolved systems are weaker than the simulation prediction \cite{Andreon2023, Gobat2019,mascolo2023,Marrewijk2024}. 
The simulated $Y$ consistently falls below the self-similar expectation since $z\gtrsim3$, suggesting a cooler ICM in protoclusters at these early epochs \cite{Rohr2025}, in contrast to a hotter ICM inferred from our observed tSZ decrement, with a 6.4$\sigma$ deviation. 
Although the gas temperature can be high at such an early epoch, the hot-gas fractions of these high-redshift cluster progenitors (Extended Data Fig.\,\ref{fig:sim_icm}) remains too small to explain our observed thermal energy excess (Extended Data Fig.\,\ref{fig:szm200}) \cite{li2023,Remus2023,Rohr2025,aljamal2025}. 

A halo mass three times larger than current observational estimates would be required for SPT2349$-$56 to match the universal $M-Y$ relation \cite{Miller2018,Hill2020,Rotermund2021}. 
The inconsistency could be even amplified if one accounts for the lower hot-gas fraction at $z>4$. 
To explain the measured tSZ strength without invoking an implausibly massive halo, intense {\it pre-heating} of the ICM must be introduced \cite{Voit2005,McCarthy2008,Hlavacek2015,Valentino2016,Henden2019,Kooistra2022,Bennett2024}. 
While cosmological hydrodynamical simulations reproduce the global cosmic star formation history, they often under-predict the high SFRs of systems like SPT2349$-$56 \cite{Bassini2020,Lim2021,Remus2023,Lim2024}. 
In principle, strong AGN feedback in SPT2349$-$56 could naturally provide the required energy to the nascent ICM  \cite{Hlavacek2015,Valentino2016,Cielo2018,Duan2020,Heckman2023,Chapman2024,Rennehan2024b,Jennings2025} (Methods). 
The elevated ambient gas pressure and density at $z>4$ confine the jets, which enhances AGN heating and reduces energy loss to expansion. 
This confinement allows a large amount of energy stored in the over-pressurized and therefore overheated ICM, producing the strong tSZ decrement seen in SPT2349$-$56 \cite{Begelman1989,Nesvadba2008,Fabian2012,Yang2016,Henden2019,Zinger2020,Bennett2024,Martin2025,flamingosz_2025,Bigwood2025,Eckert2025,Sullivan2025}. 

How common such enormous thermal reservoirs are in nascent ICM at $z\gtrsim4$ remains unknown. 
Given that SPT2349$-$56 is a unique system from a 2,500\,deg$^2$ survey, the observed strong tSZ decrement may simply reflect its own rarity, or may serve as a clue towards a potentially unrecognized but important short-lived stage of cluster assembly \cite{Fabian2012}. 
Notably, the majority of hydrodynamical simulations with state-of-the-art galaxy formation models have not predicted the extreme heating and over-pressurized hot ICM that we observe in SPT2349$-$56 \cite{Chadayammuri2021,Grayson2023,li2023,Altamura2023,Gardner2024,aljamal2025}. 
This discrepancy could point to the need for more complicated subgrid physics for AGN feedback models at high redshifts in order to match the unanticipated hot and pressurized ICM in nascent clusters \cite{Guo2008,McCarthy2008,Vogelsberger2018,Altamura2023,Weinberger2023,husko2024,Prunier2025}.

\clearpage

\begin{figure}[!h]
    \centering
    \includegraphics[width=0.98\linewidth]{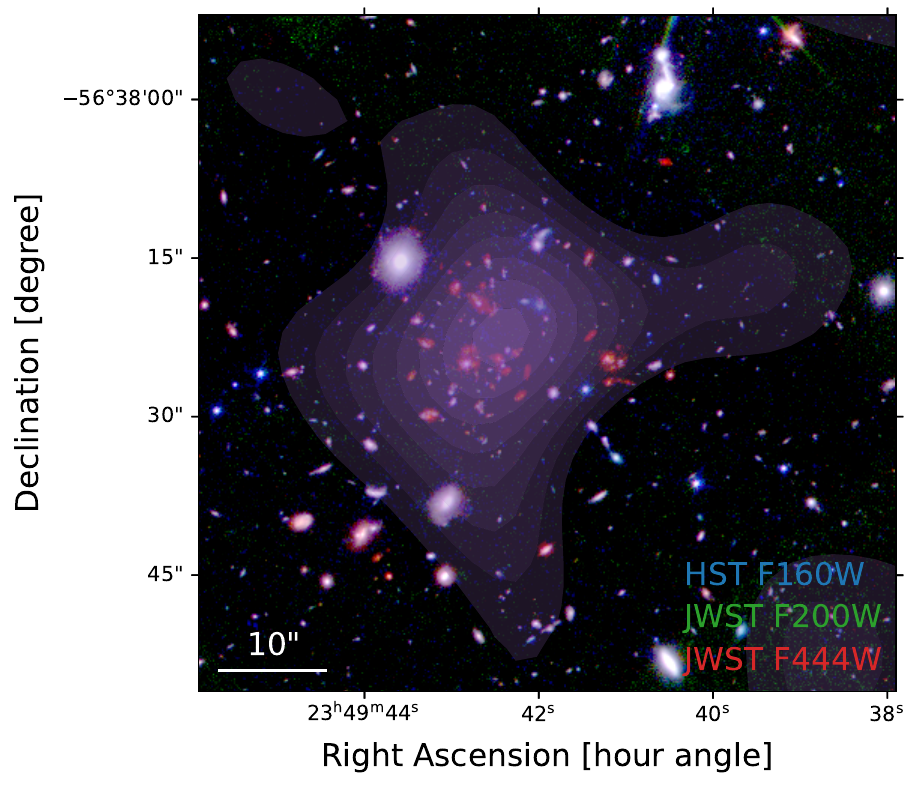}
    \caption{{\bf RGB image of the protocluster core with tSZ decrement contours.} 
    R: JWST F444W band; G: JWST F200W band; B: HST F160W band. 
    The tSZ signal (purple) was obtained after continuum source-subtraction in Fourier space and imaging done with short baselines ($uv<10k\lambda$). 
    Decrement contours are from $-1\sigma$ (8.8\,$\mu$Jy) to $-9\sigma$ (79.2\,$\mu$Jy) with steps of $-1\sigma$. 
    The decrement peak ($-8.4\sigma$) is co-spatial with the kinematic center of the protocluster system. The flux density of the tSZ signal is $-157\pm16\mu\rm Jy$ in the image plane. }
    \label{fig:jwst_sz}
\end{figure}

\clearpage

\section{Methods}\label{sec11}

Throughout the paper, we used a standard Lambda Cold Dark Matter cosmological model with $H_0\,{=}\,67.7\,\rm km\,s^{-1}\,Mpc^{-1}$ and $\Omega_m\,{=}\,0.31$ \cite{planck2015}, which corresponds to a proper angular scale of 6.9\,kpc/arcsec at $z\,{=}\,4.3$. 
$R_{200}$ and $R_{500}$ are defined as the radius where the average densities are 200 and 500 times the critical density of the Universe at the source redshift. 
$\rho_{200}$, $\rho_{500}$, $M_{200}$, $M_{500}$, $Y_{200}$, and $Y_{500}$ are the corresponding mean densities, total masses, and Compton-$Y$ parameters within these regions, respectively. 

\subsection{Observations} \label{obs}
\subsubsection{ALMA observations}

The ALMA and ACA Band-3 data of SPT2349$-$56 were obtained under four different programs from the ALMA Science Archive (Cycle 3: 2015.1.01543.T, PI: K. Lacaille; Cycle 5: 2017.1.00273.S, PI: S. Chapman; Cycle 9: 2022.1.00495.S, PI: J. Chen; Cycle 10: 2023.1.00124.S, PI: S. Chapman). 
The phase centers of these observations are similar except the Cycle 3 program, which is about 10\,arcsec to the South of the pointing centers of the other programs. 
The total on-source exposures are 23.7 and 26.9 hours for ALMA and ACA observations, respectively. 
The corresponding configurations, baseline ranges, frequency ranges, correlator modes, and on-source integration times are summarized in Extended Data Table \ref{tab:obs}.

\subsubsection{Ancillary data}

{\it Hubble Space Telescope} ({\it HST}) F160W (PID: 15701, PI: S.\,Chapman) and {\it James Webb Space Telescope} ({\it JWST}) F200W and F444W (PID: 06669, PI: S.\,Chapman) images shown in Fig.\,\ref{fig:jwst_sz} were retrieved from the Mikulski Archive for Space Telescopes (MAST) server. 
The exposure times of each filter are 8472\,s (F160W), 3264\,s (F200W), and 3264\,s (F444W), respectively. 
We followed the standard reduction procedure to process the level-1 data through {\it HST} and {\it JWST} calibration pipelines ({\tt hstcal/calwf3} and {\tt jwst/calwebb}) within the Space Telescope Environment ({\tt stenv}). 
We applied an extra de-striping step for the NIRCam images during the {\tt stage\_2} process to mitigate the influence of column stripes. 
The images were all reprojected and aligned to the F444W band using the python packages {\tt reproject} and {\tt tweakreg}. 
The detailed description of the {\it JWST} data reduction will be presented in a forthcoming paper.

\subsection{ALMA data reduction} \label{sec:alma}

For the observations carried out before 2020, we used the standard ALMA calibration script to re-calibrate the data with the latest ALMA pipeline ({\tt CASA} version=6.6.1). 
We carefully compared difference between the old and new calibrated data in their phase uncertainties, root mean square (rms) errors, and dynamical ranges of the continuum images produced from individual Execution Blocks (EBs).
In general, the calibrated data obtained from the new pipeline show a $\sim5\%$ improvement in their dynamic range. 
Therefore, we adopted all measurement sets calibrated by the new pipeline for consistency. 

We used the observatory-calibrated measurement sets for the Cycle 9 and Cycle 10 data, which were calibrated by {\tt CASA} 6.4.1 and {\tt CASA} 6.6.1, respectively. 
According to the QA2 report from the observatory, there was some potential cloud contamination in a fraction of EBs, leading to higher phase errors. 
We ran the {\tt remcloud}
pipeline for better phase calibrations and inspected the improvement of the data quality. 
Only two EBs showed $\sim20\%$ and $\sim200\%$ improvements after the {\tt remcloud} correction, which were used in this study.

Each EB was assessed by comparing the flux densities of bright sources from the {\tt CLEAN}ed image to ensure that the calibration was consistent. 
The calibration was good in general,with the amplitudes of bright sources differing by $<5\%$ at the same frequency for all EBs except two EBs, where the higher phase errors of these data introduced a loss of coherence. 
The high phase errors caused reduced amplitudes and elevated rms noises due to decorrelation, instead of amplitude calibration errors. 
In fact, the affected EBs show good agreement with other EBs after tapering, indicating that higher phase errors are dominated by the long baselines, while short baselines still keep a good coherence. 
Despite their worse qualities, these two EBs are still included for better $uv$-coverage and large-scale sensitivity. 
We used `natural' weighting for all {\tt CLEAN}ing processes in the following analysis to suppress the long-baseline contribution. 

Because the accumulated data size is over a few terabytes, some data compression was needed before imaging for a higher efficiency. 
The latest {\tt CASA} (version=6.7.0) was used for both preprocessing and imaging. 
We used the {\tt CASA} task {\tt mstransform} to compress all data to a common frequency and time width of 62.5MHz and 8s, respectively. 
This approach greatly reduces the data size and also avoids the potential bandwidth- and time-smearing effect. 
The phase center of each measurement set was then shifted to the pointing center of the Cycle 10 program through {\tt phase\_shift} command. 
Subsequently, we combined all EBs obtained from the same cycle to a single measurement set using the CASA task {\tt concat} with {\tt copypointing=False}. 
Channels with spectral lines were flagged by the task {\tt flagdata}, which are listed in Extended Data Table \ref{tab:flag}. 

Considering there are over 70 EBs with different tuning frequencies and configurations, caution is required to handle the visibility plane.
The steep spectral slope of the dust continuum emission from the protocluster galaxies can lead to baseline-dependent flux densities, owing to the frequency variations of the different baselines, which can cause artificial signals in the Fourier-plane subtraction technique. 
We used all the data to produce ultra-deep continuum images for the direct tSZ decrement imaging (`Deep' and `Deep (tapered)'), and used the subset of measurement sets with similar frequency ranges from Cycle 5 and Cycle 10 for source subtraction and tSZ measurements (`ALMA high-res', Extended Data Table \ref{tab:image}). 

The deep continuum map was generated with {\tt tclean} using line-free channels, `\texttt{natural}' weighting, `\texttt{hogbom}' deconvolver, `\texttt{mosaic}' gridder in `\texttt{mfs}' mode.
We applied a 2$''$ $uv$ taper to enhance sensitivity to extended emission. 
A two-step strategy was used for the {\tt CLEAN}ing process.
First, we {\tt CLEAN}ed without a mask down to 4$\sigma$, ensuring real emission is deconvolved while avoiding noise spikes. 
Next, we masked the high-significance 4$\sigma$ pixels and dilated them to the adjacent 2$\sigma$ pixels. 
With this robust mask, we {\tt CLEAN}ed the data to $\pm1\sigma$, which suppresses dirty-beam sidelobes and prevents over-{\tt CLEAN}ing of noise spikes. 
The rms level of the {\tt CLEAN}ed image is 1.8$\mu$Jy/beam with a $2.4''\times2.1''$ synthesized beam before tapering, and 2.1$\mu$Jy/beam with a $3.7''\times3.3''$ beam after tapering. 
In the `Deep (tapered)' image (Extended Data Fig.\,\ref{fig:clean}), a negative halo around the protocluster galaxies reaches $-4.5\sigma$, which is present in the {\tt CLEAN}ed and the residual images. 
The positive signal peaks at $116\sigma$, which is comparable to the typical dynamic range achieved in ALMA Band-3 ($\sim100\sigma$), supporting that features as faint as $-4.5\sigma$ are reliable instead of an imaging artifact. 

The tSZ signal is expected to contaminate the galaxies' continuum emission, which dominates the short spatial-frequency range. 
To produce a high-resolution continuum image for further source subtraction (`ALMA high-res'), we discarded the data with different frequency coverages (see Extended Data Table \ref{tab:image}) and excluded all baselines with $uv$-distances less than 10k$\lambda$ to suppress the extended tSZ contamination while still maintaining a good sensitivity. 
Under such a $uv$-distance selection, all signals with spatial scales above 20$''$ are greatly reduced. 
We selected the line-free channels and used the identical two-step {\tt CLEAN}ing strategy to produce the continuum map. 
With this $uv$ range-selection strategy, the continuum flux density of each source is increased by $\sim 5\%$, while the rms sensitivity only drops by $\sim1\%$. 
The synthesized beam size of the continuum image is $2.2''\times1.9''$ and the rms sensitivity is $2.1\mu$Jy/beam. 

For the direct comparison of dust continuum under different scales, we also produced two independent low-resolution images from the short baselines ($<10k\lambda$) of ALMA (`ALMA low-res') and from the ACA observations in a similar manner. 
The sensitivities of `ALMA low-res' and ACA continuum images are 11\,$\mu$Jy/beam and 24\,$\mu$Jy/beam. 
Their corresponding resolutions are $14.4''\times13.4''$ and $18.0''\times12.1''$, respectively. 

The details of each image, including the on-source exposure time, is summarized in Extended Data Table \ref{tab:image}. 
The Maximum Recoverable Scale (MRS) of each map was estimated from the corresponding spatial scale of the fifth percentile value of the ALMA baseline lengths \cite{Remijan2019,Czekala2021}. 
The quoted value could overestimate the MRS when poorly-sample and less-weighted ACA data are included, which should be treated as the upper limit. 

\subsection{Photometry measurements of continuum-emitting galaxies}\label{pho}
We used the {\tt photutils} package for the photometry measurements of continuum-emitting galaxies from the `ALMA high-res' map. 
The segmentation map was made using the {\tt detect\_sources} function with a 3$\sigma$ threshold over 10 pixels. 
{\tt deblend\_sources} (npixels=5, nlevels=32, contrast=0.001) was applied to the segmentation map, which successfully deblended source `D' and `E' (C4 and C8). 
However, due to the sparse resolution, a few sources remain blended, which were treated as single sources during the measurements. 

Subsequently, we made use of the {\tt SourceFinder} function for the source extraction. 
To avoid noise contamination, we only extracted sources with a peak pixel value $>4\sigma$. 
We recorded thee flux densities measured from their peak pixels and integrated values within standard kron apertures ({\tt k=1.4, Rmin=2.5}), with the nearby sources all masked to prevent potential contamination. 
The aperture photometry uncertainty was estimated by measuring the flux distribution in 10,000 randomly placed apertures of identical shape on the noise map. 
While the peak pixel can provide a more accurate measurement for unresolved sources, it cannot capture all emission when the source is extended. 
Therefore, we compared the flux densities obtained from these two methods and adopted the larger one as the `best' flux density (Extended Data Table \ref{tab:phot}). 
However, due to the insufficient rms sensitivity and the potential contamination from the tSZ decrement, a few confirmed protocluster members were only marginally detected ($<4\sigma$). 
For a conservative assessment of the tSZ signature, we did not include their contribution in the photometry measurement, which could cause a $\lesssim30\rm\mu Jy$ underestimation of the total dust continuum within the core. 

We also performed the continuum measurements using two independent datasets, the `ALMA low-res' and `ACA' maps, using identical methods. 
The continuum flux densities in both images are lower than the integrated values from individual galaxies measured from the high-resolution image (see Extended Data Table \ref{tab:phot}). 
The fainter emission is contrary to the general expectation from tapering, which usually shows higher fluxes for bright sources at the phase center \cite{Fujimoto2016,Fujimoto2024}. 
If the deviations are purely caused by the tSZ contamination, the signal of the tSZ decrement should be at least 131$\pm32\,\mu$Jy. 
This 4.2$\sigma$ deficit confirms the reliability of the 2.3$\sigma$ difference seen in shallower data in previous work, which was speculated to be due to the tSZ contamination by \citet{Zhou2025}. 
Because the tSZ decrement could be more extended than the size defined by continuum emission and the contribution of the faint sources was ignored, the intrinsic strength of the tSZ could be underestimated in this way. 

\subsection{Validation of the decrement with ALMA simulations}\label{sim}

To test the reliability of the potential tSZ signatures, we conducted interferometric simulations using the {\tt CASA} function {\tt simalma}. 
We used the Band-7 catalog \cite{Hill2020} as a prior and scaled the flux densities by $S_{\rm 3mm,total}/S_{\rm 850\mu m,total}$ to obtain the 3\,mm flux density of each emissive source. 
As sensitivity can be underestimated by {\tt simalma}, we tweaked the scaling factor of the telescope time in each configuration until the rms agreed with the actual value ($\Delta \rm rms/rms<5\%$). 
For a more realistic simulation, we also injected continuum sources on the jackknifed measurement sets, which were produced by inverting half of the visibilities to cancel the astrophysical signals. 

Extended Data Fig.\,\ref{fig:sim} shows the comparison between the simulations and the actual continuum image on the same scale. 
While the previous section reinforces the evidence of a tSZ decrement already seen as a negative halo in the real image, we also considered the possibility that the evidence of tSZ effect was caused by an interferometric artifact from combining different ALMA configurations. 
Although the issue of the negative sidelobes is greatly suppressed in such deep observations, imperfect {\tt CLEAN}ing can be still expected, which may cause a similar negative halo. 
However, the coherent negative halo does not exist in the simulated image, where the minimum pixel value in the core region is only 4.5\,$\mu\rm Jy$, $1.7\sigma$ less than the decrement peak (7.6\,$\mu\rm Jy$) in the actual continuum image.
Furthermore, we used the same $uv$-range ($uv<10k\lambda$) to produce the low-resolution mock image and performed the photometry measurements in an identical way. 
The continuum flux density from the low-resolution simulated map is consistent with the integrated value from the high-resolution image, which is contrary to the $20\%$ drop seen in the real data. 
Therefore, a potential artifact is unlikely to result in the current evidence of the tSZ decrement. 

\subsection{Continuum subtraction and tSZ measurement}\label{sub}
To explore the morphology and actual strength of the negative signal free from contamination from dusty galaxies, we further subtracted the known continuum sources in the $uv$-plane. 
We used the {\tt CASA} class {\tt componentlist} to construct catalogs of continuum sources as point-like for the ACA and ALMA visibilities, while considering the primary beam attenuation. 
Next, we Fourier transformed the source list from the image plane to the $uv$-plane with the {\tt ft} function and recorded the expected visibilities in the model column of each measurement set, which was then subtracted using the {\tt uvsub} command from the calibrated visibilities. 

With the continuum-subtracted data, we deconvolved the image created by the visibilities with $uv$ distances $<10k\lambda$ and generated an initial {\tt CLEAN}ed image with the `\texttt{hogbom}' deconvolver down to $4\sigma$ without the {\tt CLEAN} mask. 
We masked pixels with an absolute value $>4\sigma$ and expanded to the adjacent $>2\sigma$ pixels. 
Then, we used the multi-scale deconvolver with scales of [0, 2, 4, 8] pixels and further {\tt CLEAN}ed the data down to $1\sigma$, resulting in an rms sensitivity of 8.8$\mu$Jy/beam with a synthesized beam of $13.6''\times14.7''$. 
As shown in Extended Data Fig.\,\ref{fig:SZ}, a strong decrement occurs at the center of the protocluster core \cite{Hill2020}, with the peak $\rm \Delta(R.A., Dec.)=(-7.2'',2.1'')$ offset from the radio-loud AGN C. 

The curve-of-growth analysis (Extended Data Fig.\,\ref{fig:fwhm}) indicates that the full width at half maximum (FWHM) of the decrement is $\rm FWHM_{\rm SZ}\approx14''$ (100\,kpc) with the signal extending out to $R_{\rm SZ}\approx20''$ (140\,kpc) after correcting for beam broadening. 
This corresponds to an angular scale of $2R_{\rm SZ}\sim40''\approx 5\,k\lambda$. 
The tSZ decrement is $-100\pm12\,\rm\mu Jy$ within R$_{500}$ (80\,kpc, \cite{Chapman2024}) and the total signal is $-157\pm16\rm\,\mu Jy$. 
After correcting for projection effects \cite{Arnaud2010}, we calculated the Compton-$Y$ parameter through the formula \cite{Sunyaev1972,Carlstrom2002,Mroczkowski2019}
\begin{equation}
    \Delta I_\nu = \frac{x^4e^x}{(e^x-1)^2}\left(x\frac{e^x+1}{e^x-1}-4\right)I_0y=g(x)I_0y,
\end{equation}
where the CMB intensity $I_0=270.33\rm\,MJy/sr$, $\Delta I_\nu$ is the distortion signal in the CMB map (i.e., decrement intensity in the continuum map), and the dimensionless frequency $x=h\nu/kT_{\rm CMB}=\nu/56.81\rm\,GHz\approx1.65$. 
The Compton-$y$ map is shown in Extended Data Fig.
We obtained mean Compton-$y$ parameters of $(10.2\pm1.3)\times 10^{-6}$ and $(5.6\pm0.8)\times 10^{-6}$ for within $R_{500}$ and within $R_{\rm SZ}$, respectively. 
The integrated tSZ signal (Compton-$Y$) can be calculated as \cite{Mroczkowski2019}
\begin{equation}
    Y_{\rm SZ} = \int y\,d\Omega=\frac{1}{g(x)}\frac{1}{I_0}\int\Delta I_\nu d\Omega=\frac{S_\nu}{g(x)I_0},
\end{equation} 
where $\Omega$ is the solid angle of the selected region, and $S_{\nu}$ is the flux density in the same region. 
This leads to Compton-$Y$ parameters of $Y_{500}=(1.3\pm0.2)\times10^{-6}\rm\,arcmin^2$ and $Y_{\rm total}=(2.0\pm0.2)\times10^{-6}\rm\,arcmin^2$.

However, considering the large radius of the tSZ decrement, analysis in the $uv$-plane is needed to evaluate if the signal is limited by insufficient short baselines. 
We used the {\tt uvplot} package to export all associated data to a single $uv$ table. 
The corresponding visibilities were then shifted to the position of the decrement peak. 
Extended Data Fig.\,\ref{fig:uvplot} shows the visibilities as a function of $uv$ distance, which were averaged over every 3$k\lambda$. 
The uncertainties were estimated through the corresponding weight of each visibility. 
The positive signal of $\sim30\rm\,\mu Jy$ at 10$k\lambda$ is likely from the conservative continuum subtraction strategy, which is consistent with the estimation of the total under-subtraction signal $\sim30\rm\,\mu Jy$. 
Extended Data Fig.\,\ref{fig:uvplot} shows that the amplitude of the negative signal profile peaks at the shortest $uv$ distance of $\sim3k\lambda$ with a large uncertainty, indicating that the current data have less capability in recovering the signal at this scale. 
This suggests that the signal can approach or even go beyond the MRS and its  angular scale should be $\lesssim6k\lambda$ ($\gtrsim35''$ or 240\,kpc), which is consistent with the analysis conducted in the image plane. 
Because of the large uncertainty in the shortest $uv$ distance and the primary beam attenuation for such a large scale signal, the intrinsic tSZ signal cannot be directly measured in the $uv$-plane and a model-dependent value is out of the scope of this letter, which is planned for a future paper. 

\subsection{Comparison with other tSZ systems}\label{secA} 
To place our tSZ detection in a broader context, we compiled a sample of systems with tSZ detections, including low-redshift galaxy clusters with weak-lensing mass estimates \cite{Marrone2012,Nagarajan2019,Bocquet2019} and high-redshift systems ($z\gtrsim2$) with tSZ detections \cite{Mantz2014,Mantz2018,Gobat2019,mascolo2023,Marrewijk2024}. 
Since the reported tSZ decrements are derived from different methods and expressed in different units, we adopted a universal pressure profile (`arnaud10') to convert all results to the spherical $Y_{500}$ integrated within $R_{500}$ in units of $\rm Mpc^2$ for consistency \cite{Arnaud2010}. 
For high-redshift systems, we also used the `diemer19' halo concentration profile to convert the virial masses estimated from velocity dispersions to $M_{500}$ values \cite{Diemer2018,Diemer2019}, as most targets lack weak-lensing mass calibration. 

The ICM is known to evolve self-similarly with redshift. 
To compare our detected tSZ signal to the expected $M-Y$ relation, we removed the redshift dependence by factoring out the $E(z)^{2/3}$ scaling for each system. 
As shown in Extended Data Fig.\,\ref{fig:summary}, most low-redshift systems follow the relation $Y_{500}\propto M_{500}^{1.79}$ \cite{plancksz}. 
However, with the exception of Cl1449, high-redshift systems show a larger scatter and lie below the values predicted by the universal relation, in contrast to the tighter relation at $z\sim2$ predicted by current simulations \cite{aljamal2025}. 
This discrepancy can be explained by the presence of cooler, unvirialized ICM in the early stages of cluster formation \cite{Rohr2025}. 
Nevertheless, the tSZ decrement we observed is a factor of five larger than the predicted value, suggesting significantly higher gas pressure in SPT2349$-$56 at a much earlier epoch. 

\subsection{Thermal energy budget of the nascent ICM} \label{ori}
To examine if the decrement can be caused by gravity, we calculated the thermal energy required by the measured tSZ signal and possible gravitational energy of the virialized gas. 

We estimated the total thermal energy using the equation \cite{Spacek2016}
\begin{equation}
E_{\rm therm} =2.9\frac{m_ec^2}{\sigma_T}l_{\mathrm{ang}}^2\int y\,d\Omega=2.9 \times 10^{60} \mathrm{erg}\left(\frac{l_{\mathrm{ang}}}{\mathrm{Gpc}}\right)^2 \frac{Y_{\rm SZ}}{10^{-6} \operatorname{arcmin}^2},
\end{equation}
where the angular diameter distance to the source $l_{\rm ang}=1.42\rm\,Gpc$. 
We obtained a total thermal energy of $E_{\rm therm,total} = (11.8\pm1.2)\times10^{60}\rm\,erg$ and $E_{\rm therm,500} = (7.5\pm0.9)\times10^{60}\rm\,erg$ within the $R_{500}$ radius.  

To assess if this thermal energy can be matched to the energy of a thermal-equilibrium ICM, we need to explore its constraint on halo mass ($M_{200}$), ICM temperature ($T_{\rm ICM}$), and ICM fraction ($f_{\rm ICM}=\rm M_{\rm ICM}/M_{200}$). The halo mass $M_{200}$ can be obtained by 
\begin{equation}
    \begin{aligned}
        M_{200}&=\frac{2E_{\rm therm}\mu m_p}{3f_{\rm ICM}k_B} \frac{1}{T_{\rm ICM}},
    \end{aligned}
\end{equation}
where the molecular weight $\mu\approx0.6$, proton mass $m_p\approx1.67\times10^{-24}\rm\,g$, and $f_{\rm ICM}$ is the mass ratio between the dark matter halo and the hot ICM. 
Here we assumed that the thermal energy within $R_{200}$ radius is $E_{\rm therm,200}\approx E_{\rm therm, total}$. 
This is a good approximation because $R_{200}=18.5''\times(M_{200}/10^{13}M_\odot)^{1/3}$, which is comparable to the tSZ radius $R_{\rm SZ} \approx20''$.  

Next, we calculated the gas temperature for a virialized ICM. 
For simplicity, we used the mean virial temperature $\langle T_{200}\rangle$ within $R_{200}$ given by \citet{Voit2005},
\begin{equation}\label{virial}
k_{\mathrm{B}} T_{\rm vir} \approx k_{\mathrm{B}} \langle T_{200}\rangle=\frac{G M_{200} \mu m_p}{2 R_{200}},
\end{equation}
where the gravitational constant $G\approx6.67\times10^{-11}\rm\,m^3\,kg^{-1}\,s^{-2}$. 
The mass $M_{200}$ and radius $R_{200}$ of the halo have the following relation 
\begin{equation} \label{density}
    M_{200}=\frac{4}{3}\pi R_{200}^3 \rho_{200}.
\end{equation} 
Combining Eq.\,\ref{virial} and Eq.\,\ref{density}, we can obtain the relation between $M_{200}$ and $T_{\rm vir}$ for a virialized halo
\begin{equation}
    M_{200} = \left(\frac{6}{\pi\rho_{200}}\right)^{1/2} \left(\frac{k_B}{G\mu m_p}\right)^{3/2}T_{\rm vir}^{3/2}.
\end{equation}
If the system is virialized with a halo mass $M_{200}=(9\pm5)\times10^{12}\,M_\odot$ \cite{Hill2020}, we can derive a total thermal energy of $(f_{\rm ICM}/0.02)\times (8.5\pm3.2)\times10^{59}$\,erg, which is negligible compared to the ICM thermal energy. 
Even assuming an extremely abundant ICM $f_{\rm ICM}=6\%$ \cite{Chapman2024}, the derived thermal energy is only $(2.6\pm1.0)\times10^{60}\,\rm erg$, $\sim20$\% of the value inferred by the observed decrement.
Under this condition, the estimated thermal energy corresponds to a Compton-$Y$ parameter of $Y=(4.4\pm0.4)\times10^{-7}\mathrm{arcmin^2}$ or a distortion signal of $34\pm4\,\mathrm{\mu Jy}$. 
For a $9\times10^{12}\,M_\odot$ halo, it is unlikely that gravity serves as the dominant energy source for the observed tSZ signal. 

We then explored the possible halo masses and ICM temperatures allowed by the tSZ decrement. 
The mass of the hot ICM is limited by the available baryonic budget in the halo. 
We have a lower limit of the halo mass of 
\begin{equation}
    M_{200}\geq \frac{1}{f_b-f_{\rm ICM}}(M_{\rm mol} + M_*),
\end{equation}
where $f_b$ is the baryonic fraction of the protocluster. 
In the following calculation, we adopted the universal baryonic fraction $f_b=0.155$. 
For a conservative estimate, we used $r_{41}=0.60$ and $\alpha_{\rm CO}=1\rm\,M_\odot/(K\,km\,s^{-1}\,pc^2)$ to calculate the molecular gas mass $M_{\rm mol}$ from the CO(4--3) luminosity of SPT2349$-$56 \cite{Zhou2025}. 
As a result, $M_{\rm mol}$ and $M_*$ of SPT2349$-$56 are $4.9\,{\times}\, 10^{11}\rm\,M_\odot$ 
and $6.3\times10^{11}\rm\,M_\odot$, respectively (Pillai et al.\,in prep). 

The parameter space constrained by the measured tSZ decrement and the available baryons is shown in Extended Data Fig.\,\ref{fig:szm200}. 
Within the possible $M_{\rm ICM}$ range and without assuming an extreme $f_{\rm ICM}$, either a more massive virialized halo or a super-virialized ICM gas is necessary to match the observed tSZ decrement. 

In the case that the thermal energy is under thermal equilibrium in an extremely massive halo, we can obtain 
\begin{equation}
    M_{200} \approx (4.4\pm0.5)\times 10^{13}M_\odot \times \left(\frac{0.02}{f_{\rm ICM}}\right)^{3/5}. 
\end{equation}
Assuming a regular ICM fraction of $f_{\rm ICM}\lesssim 0.02$ at $z>4$, the expected halo mass will be more than five times the halo mass derived from velocity dispersion \cite{Hill2020}, which is unlikely given the compactness of the system. 

\subsection{Redshift evolution of ICM in TNG-Cluster simulations}\label{tng}
We compared our measured tSZ signal with predictions from the TNG-Cluster zoom-in simulations of massive galaxy clusters \citep{Nelson19a, Nelson24}. 
In brief, TNG-Cluster comprises a suite of 352 high-resolution re-simulations of massive clusters selected from a parent dark matter-only simulation with a box size of $\sim1$~Gpc. 
The sample includes all clusters at $z = 0$ with $M_{200} > 10^{15} \, M_{\odot}$, along with a representative subset of lower-mass systems spanning $M_{200} \sim 10^{14.3}$–$10^{15.0} \, M_{\odot}$. 
As with the IllustrisTNG framework, TNG-Cluster employs the \texttt{AREPO} moving-mesh magnetohydrodynamics code \citep{Springel10, Weinberger20}, together with the TNG galaxy formation model \citep{Weinberger17, Pillepich18a}, which self-consistently follows the evolution of gas, stars, supermassive black holes, chemical enrichment, and feedback from both stellar and AGN sources.

For this analysis, we used the TNG-Cluster predictions to compare the thermal component of the gas to that inferred for the SPT2349$-$56 protocluster core. 
Specifically, we computed the Compton-$y$ parameter following the prescriptions of \citet{Roncarelli07, Kay12, McCarthy14, Nelson24, Bigwood2025}, which are virtually equivalent. 
For each gas cell, the $y$ parameter ($\Upsilon_i$) was computed as 
\begin{equation}
\Upsilon_i = \frac{k_B \sigma_T}{m_e c^2} N_{e, i}  T_i,
\label{eq:Compton-y-i}
\end{equation}
where $k_B$ is Boltzmann's constant, $\sigma_T$ the Thomson scattering cross-section, $m_e$ the electron mass, and $c$ the speed of light. 
Here, $N_{e, i}$ and $T_i$ represent the number of electrons and the temperature of the $i$-th gas cell, respectively. 
The number of electrons was computed as $N_{e,i} = n_{e,i}  m_i / \rho_i$, where $n_{e,i}$ is the electron number density, $m_i$ is the gas cell mass, and $\rho_i$ is the gas density.

In practice, we derived $\Upsilon_i$ for each gas cell using \texttt{InternalEnergy} ($U$), \texttt{ElectronAbundance} ($X_e$), and \texttt{Masses} ($m_i$) from the TNG-Cluster outputs. The temperature of each gas cell was then calculated as 
\begin{equation}
T_i = \frac{U (1 - \gamma) \mu}{k_B},
\end{equation}
where $\gamma = 5/3$ is the adiabatic index and $\mu$ is the mean molecular weight, given by:
\begin{equation}
\mu = \frac{4}{1 + 3X_H + 4X_H X_e} m_p,
\end{equation}
with $X_H$ the hydrogen mass fraction (assumed to be $X_H = 0.76$) and $m_p$ the proton mass. The number of electrons was computed as:
\begin{equation}
N_{e,i} = \frac{n_{e,i} m_i}{\rho_i} = \frac{X_e X_H m_i}{m_p}.
\end{equation}

To compute $N_{e,i}$ and $T_i$ for each gas cell, we adopted a fixed hydrogen mass fraction of $X_H = 0.76$, following common assumptions in the literature \citep{Kay12, Nelson24}. 
Although one might expect a higher hydrogen fraction at earlier epochs, we verified that the median hydrogen abundance in gas cells associated with the TNG-Cluster halo at $z \sim 5$ remains close to $X_H \sim 0.76$. 

The integrated tSZ signal for a given halo, expressed as $Y_{500} D_A^2$, was calculated by summing the $\Upsilon_i$ contributions (from Equation~\ref{eq:Compton-y-i}) of all gas cells within a spherical aperture of radius $R_{500}$. 
We then tracked the redshift evolution of this tSZ proxy by evaluating $Y_{500}$ for the main progenitors of all 352 clusters identified at $z = 0$ within the TNG-Cluster suite. 
We traced $Y_{500}$ of each cluster for every four snapshots and calculated the 16th, 50th, and 84th percentile values in each redshift bin. 
The evolution of the fraction of hot gas $>10^7\rm\,K$ within the $R_{500}$ radius ($f_{\rm hot}=M_{\rm hot,500} /M_{\rm 500}$) was also recorded to investigate the simulation expectation for the hot gas at earlier epochs. 
The $Y_{500}$ values were scaled by $M_{500}^{1.79}$ to removed the mass dependency \cite{plancksz}. 

We show the 16th-50th-84th percentile values of the mass-scaled $Y_{500}$ and $f_{\rm hot}$ at $0\leq z\leq5$ for each redshift bin in Fig.\,\ref{fig:z_evo} and Extended Data Fig.\,\ref{fig:sim_icm}, respectively. 
As indicated by Fig.\,\ref{fig:z_evo}, the tSZ decrement observed in SPT2349$-$56 is also more than five times the value predicted by TNG-Cluster. 
Without assuming a very massive halo (Extended Data Fig.\,\ref{fig:szm200}), the lower hot-gas fraction or a cooler ICM in TNG-Cluster is likely the reason of this discrepancy. 

\subsection{Possible energy injection from non-gravitational processes} \label{agn}
In addition to energy from gravitational processes, AGN and star-formation activities are likely to provide the additional energy injection needed to explain the large thermal reservoir inferred by the observed tSZ decrement. 
Given that kinetic energy is the dominant source besides gravitational energy for large-scale heating \cite{Fabian2012}, we only consider its contribution to the thermal energy of the nascent ICM in SPT2349$-$56 for simplicity.  

\subsubsection{Kinetic-mode AGN feedback}

Previous studies indicate that AGN activity is significantly enhanced in SPT2349$-$56 \cite{Chapman2024,Vito2024}. 
At least three protocluster members show strong radio excess, with a total rest-frame 1.4\,GHz power of $L_{\rm 1.4\,GHz}=(2.2\pm0.3)\,{\times}\,10^{26}\,\mathrm{W/Hz}$ (Chapman et al. in prep), of which one is also luminous in X-rays \cite{Vito2024}. 
From their modest radio luminosities, \citet{Chapman2024} speculates that the detected radio AGN can be fueled by hot gas in radio-mode instead of radiative-efficient accretion through recent mergers, which can provide strong kinetic feedback with substantial energy injection on the nascent ICM in the protocluster. 
We used the correlation between the cavity power ($P_{\rm cav}$) and radio luminosity at 1.4\,GHz ($L_{\rm 1.4\,GHz}$) to estimate the kinetic power from radio luminosity alone, which can be described using \cite{Heckman2014}
\begin{equation}
    P_{\rm cav}=7\times10^{43}{\rm \,erg/s}\times f_{\rm cav}\left(\frac{L_{\rm 1.4\,GHz}}{10^{25}{\rm W\,Hz^{-1}}}\right)^{0.68} ,
\end{equation}
where the enthalpy factor $f_{\rm cav}=4$ for relativistic plasma. 
Shocks induced by the radio jet can cause additional heating, which could imply a higher $f_{\rm cav}>4$ \cite{Nusser2006,Jennings2025}. 
This scaling relation yields a kinetic power of $\dot{E}_{\rm kin, radio}=(2.3\pm0.3)\times10^{45}{\rm\,erg/s}\times(f_{\rm cav}/4)$. 

\subsubsection{Radiative-mode AGN feedback}
On the other hand, the protocluster galaxy `A' (or `C1') is luminous in X-ray with an AGN luminosity of $L_{\rm AGN}=(1.9\pm0.7)\times10^{47}\rm\,erg/s$ \cite{Vito2024}, which can also power large-scale outflows through radiation pressure when the outflow exceeds the local escape velocity \cite{Fabian2012,Rennehan2024b}.
We can use the tight correlation between AGN luminosity and outflow rate to estimate the kinetic power from the radiation pressure: 
\begin{equation}
    \dot{E}_{\rm kin,rad} = f_{\rm rad}\times L_{\rm AGN},
\end{equation}
where the coupling efficiency of radiative-mode feedback $f_{\rm rad}\approx0.5\%$ \cite{Heckman2023,Kondapally2023}. 
We note that energy can be injected to the forming ICM only when outflows overcome the potential well of the host galaxy. 
The corresponding escaping power is 
\begin{equation}
    \dot{E}_{\rm esc,rad}=f_{\rm esc,rad}\dot{E}_{\rm kin, rad}=\left[1-\left(\frac{v_{\rm esc}}{v_{\rm out}}\right)^2\right]\dot{E}_{\rm kin,rad},
\end{equation}
where $f_{\rm esc,rad}$ is the escape efficiency for the radiation-driven outflow, $v_{\rm esc}=\sqrt{2GM/R}$ is the escape velocity from the host galaxy, and $v_{\rm out}$ is the velocity of AGN outflow driven by the radiative feedback. 
With the assumed AGN-outflow velocity of $v_{\rm out}\approx1000\rm\,km/s$, estimated dynamical mass $M_{\rm dyn}\approx2.7\times10^{11}\rm\,M_\odot$, and galaxy radius $R\approx6\rm\,kpc$ \cite{Hill2020,Venkateshwaran2024}, the escape efficiency is $f_{\rm esc,rad}\approx0.6$. 
The escaping kinetic power from `A' is $\dot{E}_{\rm sec,rad}=(0.6\pm0.2)\times10^{45}{\rm\,erg/s}\times(f_{\rm esc,rad}/0.6)\times(f_{\rm rad}/0.005)$. 

\subsubsection{Star formation}
We next consider energy injection from star formation activity. 
The kinetic power from star formation-driven outflows is given by \cite{Spilker2020,Chapman2024}
\begin{equation}
   \dot{E}_{\rm esc,SF}\approx \frac{1}{2}\dot{M}_{\rm out,SF} v^2_{\rm out,SF} f_{\rm esc,SF},
\end{equation}
where $f_{\rm esc,SF}$ is the escape efficiency of an outflow and the mass outflow rate $\dot M_{\rm out, SF}$ is related to the SFR by the mass-loading factor $\eta=\dot M_{\rm out,SF}/\rm SFR\approx1$. 
Given an SFR of $\sim(5000\pm600)\rm\, M_\odot/yr$ in the protocluster core \cite{Miller2018,Hill2020,Chapman2024}, the kinetic power from star formation is $\dot{E}_{\rm esc,SF}=(0.8\pm0.1)\times10^{44}{\rm \,erg/s}\times (v_{\rm out,SF}/500{\rm\, km\,s^{-1}})^2\times(f_{\rm esc,SF}/0.2)$, which is less than $5\%$ of the kinetic power from AGN feedback. 
We therefore conclude that energy injection from star formation can be safely neglected.

\subsubsection{Total energy injection and thermal coupling efficiency}
Assuming $f_{\rm cav}=4, f_{\rm esc, rad}=0.6,$ and $f_{\rm rad}=0.005$, we can estimate the total energy input: 
\begin{equation}
\begin{aligned}
        \Delta E_{\rm inject} &\approx \dot{E}_{\rm kin,radio} t_{\rm radio} +  \dot{E}_{\rm esc,rad} t_{\rm rad}\\
        &= \left[(7.3\pm0.9)\times  \left(\frac{t_{\rm radio}}{100\rm\,Myr}\right) + (1.8\pm0.7)\times \left(\frac{t_{\rm rad}}{100\rm\,Myr}\right) \right]\times10^{60} \rm\, erg. 
\end{aligned}
\end{equation}
Adopting $t_{\rm AGN}=100\,\rm Myr$ for  AGN activities, we can obtain a total energy injection of $(9.1\pm1.1)\times10^{60}\rm\,erg$ to be stored in the nascent ICM. 

According to thermodynamics and the ideal gas law, the kinetic energy is known to have two effects. 
The injected energy can lead to a system expansion against the ambient pressure or an increase in the thermal energy of the ICM, which can be expressed as 
\begin{equation}
    \Delta E_{\rm inject} \propto \int P{\rm d}V+\int V{\rm d}p =\Delta E_{\rm therm}+W_{\rm expan}=\frac{\Delta E_{\rm therm}}{f_{\rm therm}},
\end{equation}
where $\Delta E_{\rm therm}$ is the thermal energy change due to the pressure increase, $W_{\rm expan}$ is the expansion work, and the thermal energy coupling efficiency $f_{\rm therm}$ is defined as the fraction of kinetic power increasing the thermal energy instead of the bulk work. 
Assuming that the halo is virialized with $f_{\rm ICM}=0.02$, the thermal energy change is $\Delta E_{\rm therm} = E_{\rm therm} - E_{\rm therm, vir}=(10.9\pm1.2)\times10^{60}\rm\,erg $.

These assumptions imply a thermal coupling efficiency of $f_{\rm therm}\,{=}\,120\,{\pm}\,20\%$, indicating that our energy injection must be underestimated. 
To reproduce the observed tSZ decrement, at least one of the following parameters must exceed our fiducial values: 
\begin{itemize}
    \item The cavity enthalpy factor $f_{\rm cav}>5$ instead of 4; 
    \item The AGN active time $t_{\rm AGN}>120\,\rm Myr$ instead of 100\,Myr; 
    \item The radiative-mode coupling efficiency $f_{\rm rad}>1\%$ instead of 0.5\%; 
    \item The fraction of hot ICM $f_{\rm ICM}>6\%$ instead of 2\%; 
    \item The halo mass $M_{200}>1.7\times10^{13}\,M_\odot$ instead of $0.9\times10^{13}\,M_\odot$. 
\end{itemize}
We note that the kinetic-mode AGN feedback can supply substantially more energy, which naturally explains the inferred thermal energy. 
The radio AGN in SPT2349$-$56 exhibit steep spectral slopes ($\alpha<-1$, Chapman et al.\,in prep), indicating relatively old ages. 
\citet{Chapman2024} point out that the steep spectral slope, along with the compact size of the radio AGN, implying a potential compact steep spectrum source, which could lead to an age exceeding 500\,Myr. 
Moreover, the enthalpy factor $f_{\rm cav}$ can exceed the canonical value when the ambient gas pressure is high. 
Indeed, the ambient pressure can be sufficient enough to confine jets and enhance shock heating in some radio galaxies 
\cite{Begelman1989,Odea1998,Yamada1999,Bromberg2011,Cen2024}. 
At $z>4$, the ICM pressure are higher due to the elevated cosmic critical density, which naturally provides a strong confining pressure for the surrounding medium \cite{Boselli2022}. 
The enhanced pressure can also limit system expansion, which boosts $f_{\rm therm}$ by suppressing the $PdV$ term, resulting a large thermal energy reservoir in the nascent ICM at $z>4$. 

\clearpage

\setcounter{figure}{0}
\makeatletter 
\renewcommand{\thefigure}{\@arabic\c@figure}
\renewcommand{\thetable}{\@arabic\c@table}
\renewcommand{\figurename}{Extended Data Fig.}
\renewcommand{\tablename}{Extended Data Table}
\makeatother

\begin{figure}
    \centering
    \includegraphics[width=0.98\linewidth]{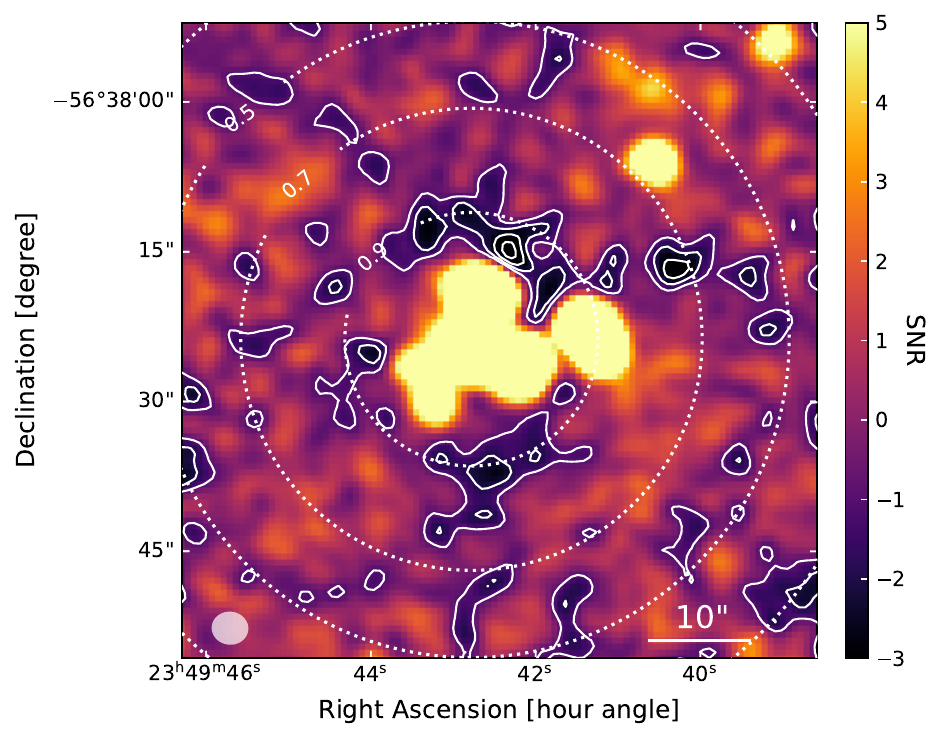}
    \caption{{\bf Tapered ALMA+ACA 3\,mm continuum map. }
    The positive signals come from dust emission of DSFGs. 
    We use the contours from $-1\sigma$ to $-4\sigma$ with steps of $-1\sigma$ to highlight the negative signals. 
    The synthesized beam is shown in the bottom-left corner and the primary beam responses of 0.3, 0.5, 0.7, and 0.9 are indicated as dotted lines. 
    A consistent negative ring is seen around the protocluster core, with a peak value at $-4.5\sigma$, suggesting the existence of extended tSZ signal. }
    \label{fig:cont}
\end{figure}

\begin{figure}
    \centering
    \includegraphics[width=0.98\linewidth]{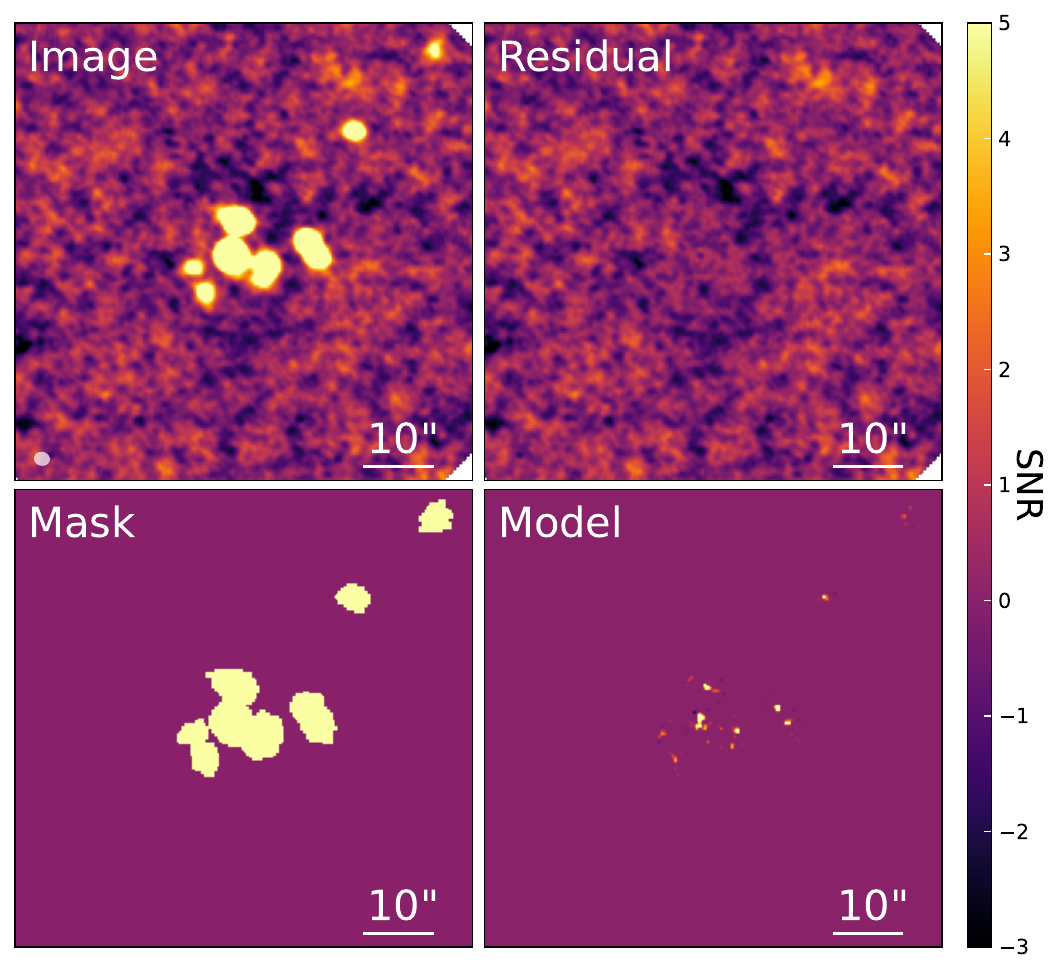}
    \caption{{\bf \texttt{\textbf{CLEAN}}ing process for the `deep' imaging.} 
    To avoid negative sidelobes from the dirty beam, we perform a deep {\tt CLEAN}ing down to the $1\sigma$ level, with the {\tt CLEAN}ing mask and model indicated in the bottom panels. 
    The synthesized beam is shown in the bottom-left corner of the upper-left panel. 
    Despite the fact that no obvious emission is left after {\tt CLEAN}ing, the negative halo persists in the residual image (top right).
    }
    \label{fig:clean}
\end{figure}
\clearpage
\begin{figure}
    \centering
    \includegraphics[width=0.995\linewidth]{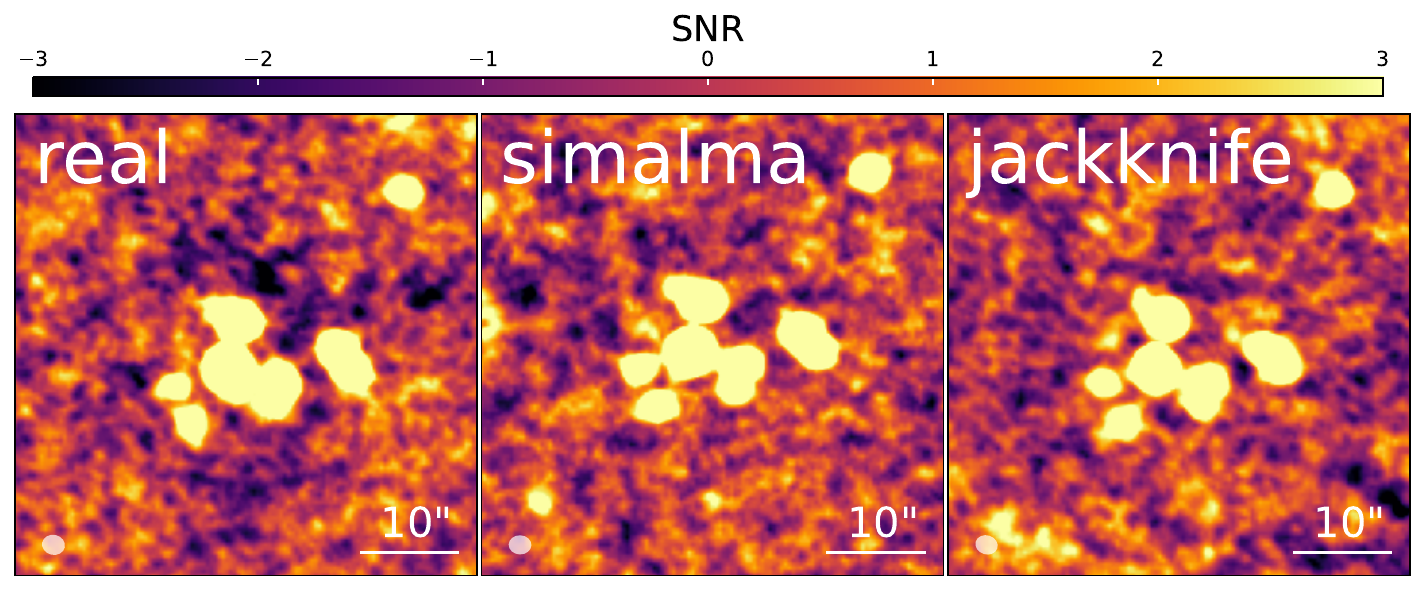}
    \caption{{\bf Comparison between the real and simulated images.} 
    We created simulated continuum images to further investigate if the negative halo can be caused by any unknown interferometric artifacts due to the complex distribution of the bright DSFGs. 
    The corresponding synthesized beam is shown in the bottom-left corner for each panel. 
    The minimum pixel value from the real map is 1.7$\sigma$ more significant than both simulated maps, and the simulated maps do not display a coherent negative ring seen in the real image. }
    \label{fig:sim}
\end{figure}
\clearpage
\begin{figure}
    \centering
    \includegraphics[width=0.98\linewidth]{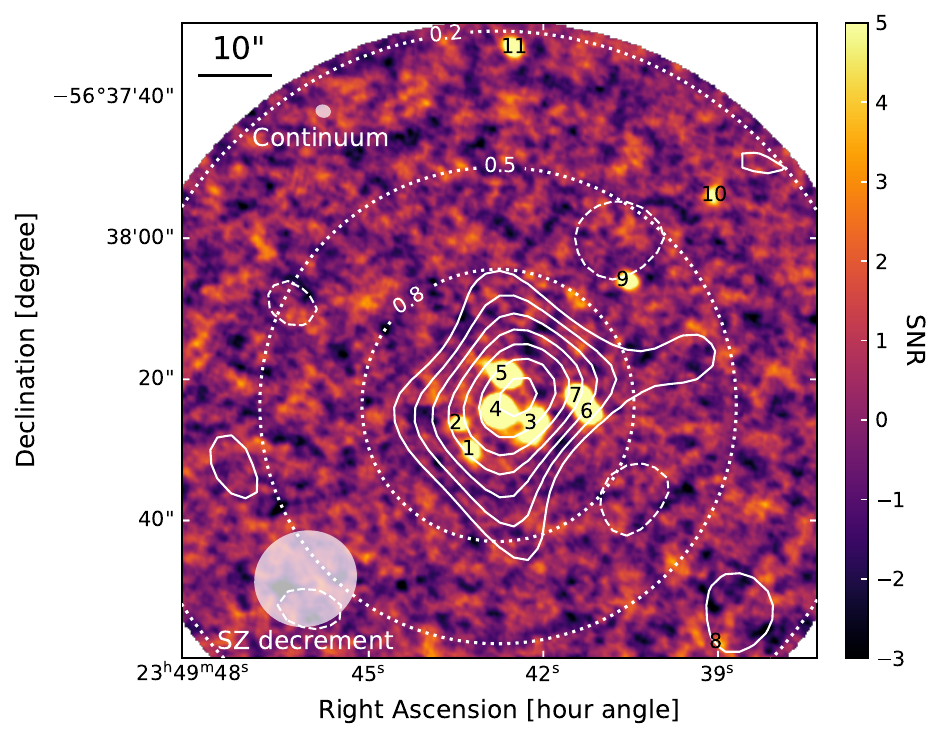}
    \caption{{\bf ALMA high-resolution continuum map with the tSZ contours.} 
    The solid contours are from $-2\sigma$ to $-8\sigma$ with steps of $-1\sigma$. 
    The dashed contours indicate regions with values above $2\sigma$. 
    The primary beam responses are indicated as dotted lines. 
    The synthesized beams of the continuum image (`ALMA high-res') and the SZ decrement (`SZ') are indicated in the upper left and the bottom-left corners, respectively. }
    \label{fig:SZ}
\end{figure}
\begin{figure}
    \centering
    \includegraphics[width=0.98\linewidth]{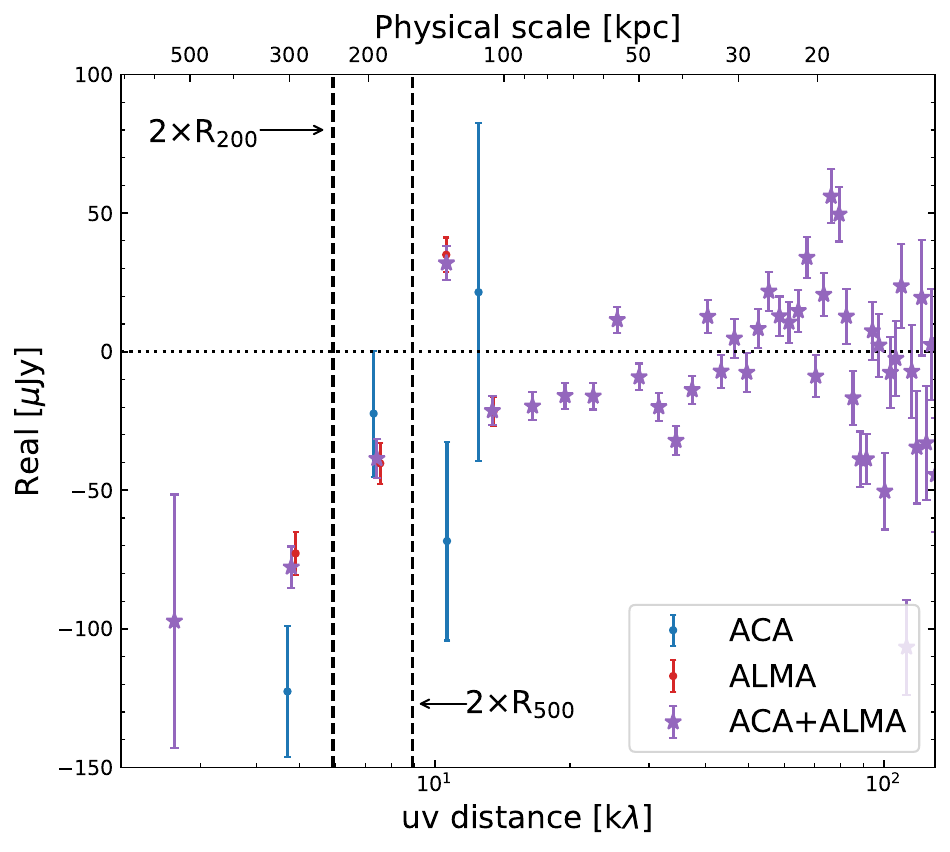}
    \caption{{\bf $\boldsymbol{uv}$ profile of the continuum subtracted data. } 
    The real part of the averaged amplitude measured from the continuum-subtracted measurement sets as a function of $uv$-distance with $uv$ bins of $3k\lambda$. 
    The negative signal becomes stronger at a shorter $uv$ distance, suggesting that the tSZ decrement can be limited by the $uv$-coverage of the ACA and ALMA observations. }
    \label{fig:uvplot}
\end{figure}
\begin{figure}
    \centering
    \includegraphics[width=0.98\linewidth]{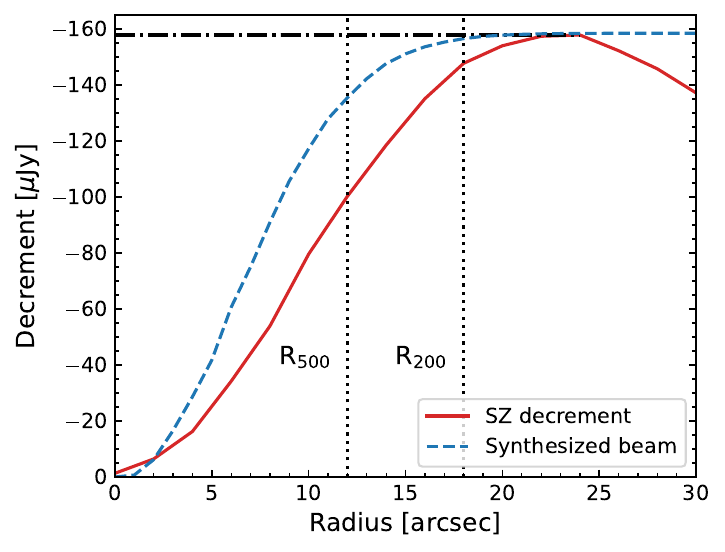}
    \caption{{\bf Curve-of-growth analysis for the tSZ decrement.} 
    The flux density of the tSZ decrement and the flux-scaled synthesized beam of the `SZ' map are indicated as the red solid and the blue dashed line, respectively. 
    The overdensity radii are also shown as black dotted lines. 
    The decrement turnaround at radii $\gtrsim25''$ may indicate potential dirty-beam sidelobes from imperfect {\tt CLEAN}ing process. }
    \label{fig:fwhm}
\end{figure}
 
\begin{figure}
    \centering
    \includegraphics[width=0.98\linewidth]{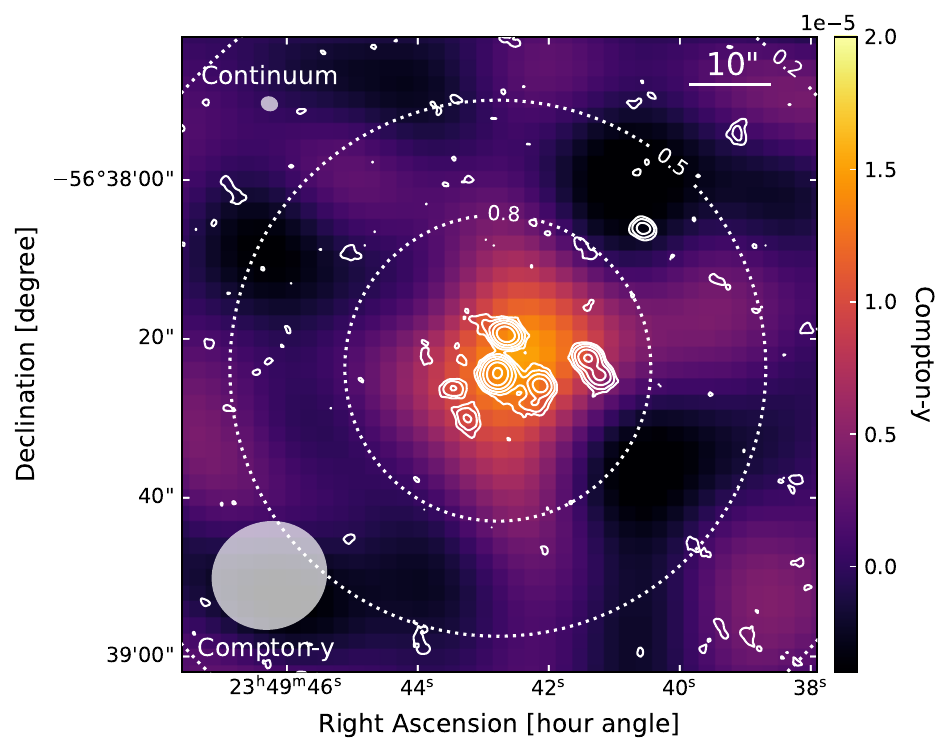}
    \caption{{\bf Compton-$\boldsymbol y$ map with dust continuum contours.} 
    The solid contours are 3\,mm continuum emission drawn at [$2.5\sigma$, $5\sigma$, $10\sigma$, $20\sigma$, $40\sigma$, $80\sigma$] from the `ALMA high-res' map. 
    The primary beam responses are indicated at dotted lines. 
    The synthesized beams of 3\,mm continuum and Compton-$y$ map are indicated in the upper-left and the bottom-left corners. }
    \label{fig:ymap}
\end{figure}

\begin{figure}[htp]
    \centering
    \includegraphics[width=0.98\linewidth]{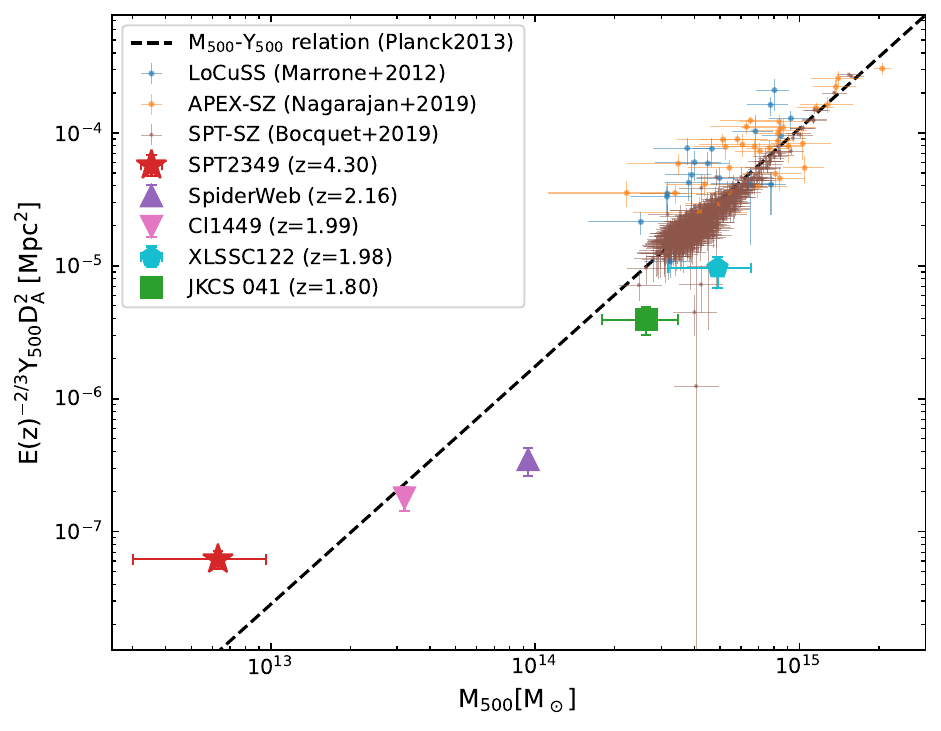}
    \caption{{\bf TSZ signal $\boldsymbol{Y_{500}}$ as a function of halo mass $\boldsymbol{M_{500}}$ within $\boldsymbol{R_{500}}$}. 
    The halo masses $M_{500}$ and Compton-$Y$ $Y_{500}$ are same as in Fig.\,\ref{fig:z_evo}. 
    The self-similar redshift evolution have been taken into account by including a factor of $E(z)^{-2/3}$. 
    The dashed line shows the universal $M_{500}-Y_{500}$ relation reported by \citet{plancksz}. 
    }
    \label{fig:summary}
\end{figure}

\begin{figure}
    \centering
    \includegraphics[width=0.98\linewidth]{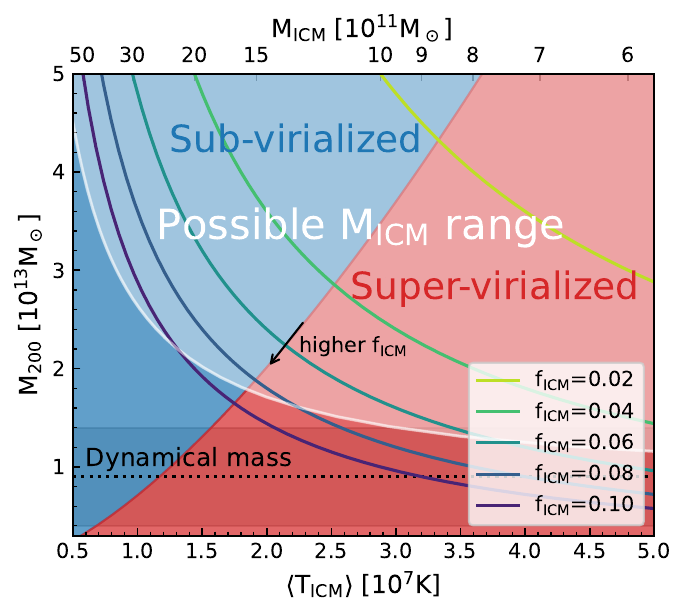}
    \caption{{\bf Halo mass as a function of ICM temperature.} 
    The corresponding ICM mass M$_{\rm ICM}$ is indicated in the top axis. 
    The sub-virialized (blue) and super-virialized (red) regions are separated by the virialized temperature for different M$_{200}$. 
    The dotted line indicates the dynamical mass of SPT2349$-$56 \cite{Hill2020}. 
    The lighter upper region is the possible M$_{\rm ICM}$ range allowed by the universal baryonic fraction $f_{\rm b}=0.155$. 
    Without assuming an extreme ICM fraction $f_{\rm ICM}$, either a more massive virialized halo or a super-virialized ICM gas is necessary to match the observed tSZ decrement.}
    \label{fig:szm200}
\end{figure}
\begin{figure}
    \centering
    \includegraphics[width=0.98\linewidth]{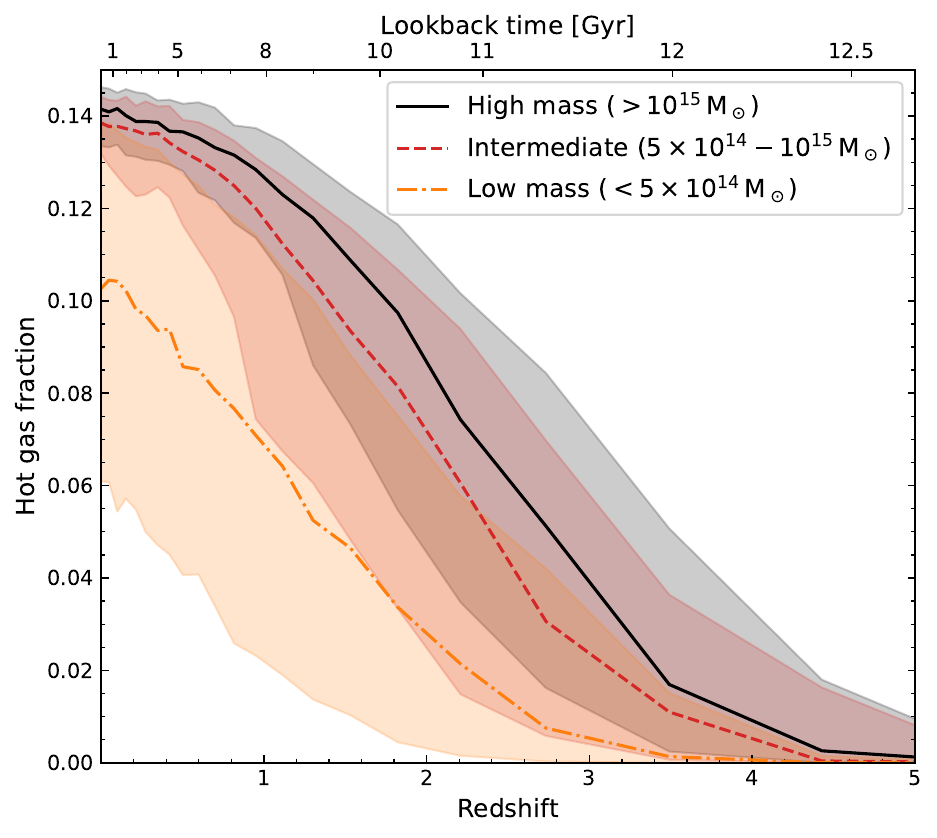}
    \caption{{\bf Cosmic evolution of the hot gas ($\boldsymbol{>10^7\rm\,K}$) fraction in galaxy clusters.} The fraction of hot ICM ($\rm M_{\rm hot\,ICM, 500}/M_{500}$) within $r_{500}$ as a function of redshift obtained from the TNG-Cluster simulation. Based on the total mass $\rm M_{200}$ at $z=0$, the clusters are placed in three mass bins, high mass ($\rm M_{200}>10^{15}\,M_\odot$, black solid line), intermediate mass ($\rm 5\times 10^{14}\,M_\odot <M_{200}<10^{15}\,M_\odot$, red dashed line), and low mass ($\rm M_{200}<5\times10^{14}\,M_\odot$, orange dash-dotted line). The corresponding shaded regions represent the values between the 16th and 84th percentiles in individual mass bins.}
    \label{fig:sim_icm}
\end{figure}

\begin{landscape}
\begin{table*}[]
    \centering
    \caption{{\bf Summary of ALMA observations used in this study.}}
    \label{tab:obs}
    \begin{tabular}{ccccc}
    \hline
    \hline
       Program ID  & Configuration & $L_{\rm base}$ & Frequency coverage & $t_{\rm source}$ \\
       && [m] & [GHz]&[min] \\
       \hline
        2023.1.00124.S & 12-m (C-3) & 15--500 & 85.9--89.9, 97.9--101.9 & 685 \\
        2023.1.00124.S & 12-m (C-4) & 15--784 & 85.9--89.9, 97.9--101.9 & 49 \\
        2023.1.00124.S & 12-m (C-2) & 15--314 & 85.9--89.9, 97.9--101.9 & 49 \\
        2023.1.00124.S & 7-m (ACA) & 9--49 & 85.9--89.9, 97.9--101.9 & 1616 \\
        2022.1.00495.S & 12-m (C-3) & 15--368 & 89.1--92.9, 101.1--104.9 & 198 \\
        2022.1.00495.S & 12-m (C-2) & 15--313 & 89.1--92.9, 101.1--104.9 & 148 \\
        2017.1.00273.S & 12-m (C43-4) & 15--784 & 86.2--89.6, 97.9--101.6 & 47  \\
        2017.1.00273.S & 12-m (C43-5) & 15--1231 & 86.2--89.6, 97.9--101.6 & 116 \\
        2017.1.00273.S & 12-m (C43-4) & 15--784& 89.8--93.7, 101.8--105.8 & 78 \\
        2015.1.01543.T & 12-m (C40-4) & 15--704& 84.3--87.9, 96.4--100.0 & 47 \\
        \hline
    \end{tabular}
\end{table*}

\begin{table*}[]
    \centering
    \caption{{\bf Frequency range of the flagged channels in each tuning.}}    \label{tab:flag}
    \begin{tabular}{cccc}
    \hline
    \hline
       Program ID  & Frequency coverage & Flagged channels & Flagged \\
       & [GHz] & [GHz] & fraction  \\
       \hline
        2023.1.00124.S & 85.9--89.9, 97.9--101.9 & 86.5--87.35 (CO), 98.5--98.85 (foreground) & 16\% \\
        2022.1.00495.S & 89.1--92.9, 101.1--104.9 & 92.4--92.9 ([CI]), 103.1--104.3 ($^{13}$CO\&C$^{18}$O) & 23\% \\
        2017.1.00273.S & 86.2--89.6, 97.9--101.6 & 86.5--87.35 (CO), 98.5--98.85 (foreground) & 16\% \\
        2017.1.00273.S & 89.8--93.7, 101.8--105.8 & 92.4--93.2 ([CI]), 103.1--104.3 ($^{13}$CO\&C$^{18}$O) & 27\% \\
        2015.1.01543.T & 84.3--87.9, 96.4--100.0 & 86.5--87.35 (CO), 98.5--98.85 (foreground) & 16\% \\
        \hline
    \end{tabular}
\end{table*}

\begin{table*}[]
    \centering
    \caption{{\bf Details of continuum maps used in this study.} }    \label{tab:image}
    \begin{tabular}{ccccccccc}
    \hline
    \hline
       Map  & Used data & Frequency coverage & $uv$-range & Taper & $t_{\rm ALMA}\,{+}\,t_{\rm ACA}$ & RMS & Beam & MRS \\
        & & [GHz] & [$k\lambda$] & & [min] & [$\mu\rm Jy/beam$] & [arcsec$^2$] & [arcsec]\\
       \hline
        Deep & all & 84.3--93.7, 96.4--105.8 & 2--400 & N/A & 1416+1616 & 1.8 & 2.4$\times$2.1 & 33\\ 
       \hline
        Deep (tapered) & all & 84.3--93.7, 96.4--105.8 & 2--400 & $2''$ & 1416+1616 & 2.1 & 3.7$\times$3.3 & 34\\ 
        \hline
        \multirow{2}{*}{ALMA high-res} & 2023.1.00124.S & 85.9--89.9, 97.9--101.9 & 10--220 & \multirow{2}{*}{N/A} & 782+0 & \multirow{2}{*}{2.1} & \multirow{2}{*}{2.2$\times$1.9} & \multirow{2}{*}{16}\\
        & 2017.1.00273.S & 86.2--89.6, 97.9--101.6 & 10--400 & &163+0\\
        \hline
        \multirow{2}{*}{ALMA low-res} & 2023.1.00124.S & 85.9--89.9, 97.9--101.9 & 4--10 & \multirow{2}{*}{N/A} & 782+0 & \multirow{2}{*}{11} & \multirow{2}{*}{14.4$\times$13.4} & \multirow{2}{*}{48}\\
        & 2017.1.00273.S & 86.2--89.6, 97.9--101.6 & 4--10 && 163+0  \\
        \hline
        ACA & 2023.1.00124.S & 85.9--89.9, 97.9--101.9 & 2--14 & N/A & 0+1616 & 24 & 18.0$\times$12.1 & $\lesssim$\,76\\
        \hline
        \multirow{2}{*}{SZ} & 2023.1.00124.S & 84.3--87.9, 96.4--100.0 & 2--10 & \multirow{2}{*}{N/A} & 782+1616 & \multirow{2}{*}{8.8} & \multirow{2}{*}{13.6$\times$14.7} & \multirow{2}{*}{$\lesssim$\,72}\\
        & 2017.1.00273.S & 86.2--89.6, 97.9--101.6 & 4--10 && 163+0\\
        \hline
    \end{tabular}
\end{table*}

\begin{table}[!tp]
\centering
\caption{{\bf 3-mm continuum source catalog.}}\label{tab:phot} 
\begin{tabular}{cccccccccc}
\hline
\hline
ID & R.A. & Dec. & Distance & S/N & PB & S$_{\rm peak}$ & S$_{\rm kron}$ & S$_{\rm best}$ & Name$^1$\\
& hh:mm:ss & dd:mm:ss & $\mathrm{arcsec}$ &  &  & $\mathrm{\mu Jy}$ & $\mathrm{\mu Jy}$ & $\mathrm{\mu Jy}$ \\
\hline
\multicolumn{9}{c}{Distance$<15''$ (ALMA high-res)} \\
1 & 23:49:43.24 & --56:38:30.06 & 9.0 & 10.7 & 0.96 & 23$\pm$2 & 35$\pm$6 & 35$\pm$6 & J \\
2 & 23:49:43.46 & --56:38:26.29 & 7.2 & 10.1 & 0.97 & 22$\pm$2 & 27$\pm$5 & 27$\pm$5 & H \\
3 & 23:49:42.17 & --56:38:26.35 & 6.5 & 30.8 & 0.98 & 66$\pm$2 & 106$\pm$6 & 106$\pm$6 & F,I,L \\
4 & 23:49:42.78 & --56:38:24.49 & 2.5 & 89.4 & 1.00 & 186$\pm$2 & 286$\pm$5 & 286$\pm$5 & B,C,G \\
5 & 23:49:42.67 & --56:38:19.37 & 2.7 & 75.2 & 0.99 & 158$\pm$2 & 185$\pm$5 & 185$\pm$5 & A, K \\
6 & 23:49:41.21 & --56:38:24.73 & 13.0 & 25.6 & 0.90 & 59$\pm$2 & 64$\pm$5 & 64$\pm$5 & E \\
7 & 23:49:41.41 & --56:38:22.49 & 11.1 & 44.4 & 0.92 & 100$\pm$2 & 107$\pm$4 & 107$\pm$4 & D \\
\hline
\multicolumn{9}{c}{Total} \\
ALMA high-res &  $\dots$  &  $\dots$  &  $\dots$  &  $\dots$  &  $\dots$  & 614$\pm$6 & 809$\pm$14 & 809$\pm$14 \\
ALMA low-res & 23:49:42.36 & --56:38:23.79 & 3.4 & 45.0 & 1.00 & 502$\pm$11 & 678$\pm$29 & 678$\pm$29 \\
ACA & 23:49:42.40 & --56:38:24.16 & 3.2 & 21.8 & 1.00 & 515$\pm$24 & 621$\pm$40 & 621$\pm$40 \\
\hline
\multicolumn{9}{c}{Distance$>15''$ (ALMA high-res)} \\
8 & 23:49:38.97 & --56:38:57.41 & 47.2 & 4.1 & 0.21 & 41$\pm$10 & 59$\pm$18 & 59$\pm$18 &  $\dots$  \\
9 & 23:49:40.56 & --56:38:06.10 & 24.1 & 14.7 & 0.65 & 47$\pm$3 & 40$\pm$5 & 47$\pm$3 & NL1 \\
10 & 23:49:39.11 & --56:37:54.02 & 41.0 & 6.6 & 0.28 & 49$\pm$7 & 67$\pm$18 & 67$\pm$18 &  $\dots$  \\
11 & 23:49:42.54 & --56:37:33.07 & 49.0 & 14.0 & 0.15 & 197$\pm$14 & 210$\pm$23 & 210$\pm$23 & N1 \\
\hline
Decrement (SZ) & 23:49:42.53 & --56:38:23.55 & 2.0 & 8.4 & 1.00 & --74$\pm$9 &  $\dots$  & --157$\pm$16 &  \\ 
\hline
\end{tabular}
\footnotetext[1]{Names of the corresponding protocluster members \cite[See Ref.][]{Miller2018}.}
\end{table}
\end{landscape}

\clearpage

\bmhead{Acknowledgments} 
We are grateful to Leslie Sage for his valuable guidance and thoughtful feedback, which greatly improved the clarity and presentation of the letter. 
We thank Arif Balbul, Gilbert Holder, Adam Mantz, Daniel Marrone, Anita Richards, Douglas Rennahan, and Bohan Yue for useful discussions on this work. 
This paper makes use of the following ALMA data: ADS/JAO.ALMA\#2015.1.01543.T, ADS/JAO.ALMA\#2017.1.00273.S, ADS/JAO.ALMA\#2022.1.00495.S, ADS/JAO.ALMA\#2023.1.00124.S. 
ALMA is a partnership of ESO (representing its member states), NSF (USA) and NINS (Japan), together with NRC (Canada), NSTC and ASIAA (Taiwan), and KASI (Republic of Korea), in cooperation with the Republic of Chile. The Joint ALMA Observatory is operated by ESO, AUI/NRAO and NAOJ. 
The National Radio Astronomy Observatory is a facility of the National Science Foundation operated under cooperative agreement by Associated Universities, Inc. 
This research used the Canadian Advanced Network For Astronomy Research (CANFAR) operated in
partnership by the Canadian Astronomy Data Centre and The Digital Research Alliance of Canada with
support from the National Research Council of Canada the Canadian Space Agency, CANARIE and the
Canadian Foundation for Innovation.
The SPT is supported by the National Science Foundation through grant PLR-1248097, with partial support through PHY-1125897, the Kavli Foundation, and the Gordon and Betty Moore Foundation grant GBMF 947. 
D.Z., S.C.C, R.H, and G.C.P.W. acknowledge support from NSERC-6740. 
M.A. is supported by FONDECYT grant number 1252054, and gratefully acknowledges support from ANID Basal Project FB210003 and ANID MILENIO NCN2024\_112.
M.S. was financially supported by Becas-ANID scholarship \#21221511, and also acknowledges support from ANID BASAL project FB210003. 

\section*{Declarations}

\begin{itemize}
\item {\bf Data availability}: 
The ALMA and {\it HST} data used in this work are publicly available on the ALMA science archive (https://almascience.nrao.edu/aq/) and MAST service (https://mast.stsci.edu). 
The {\it JWST} data were obtained as a part of program {\it JWST}-GO-06669, which will be accessible after the proprietary period.  

\item {\bf Code availability}: 
Presented results can be reproduced using the following publicly available packages: 
\texttt{astropy} \cite{astropy2022}, 
\texttt{astroquery} \cite{astroquery}, 
\texttt{CASA} \cite{CASA}, 
\texttt{colossus} \cite{Diemer2018}, 
\texttt{matplotlib} \cite{matplotlib}, 
\texttt{numpy} \cite{numpy}, 
\texttt{pandas} \cite{pandas}, 
\texttt{photutils} \cite{photutils}
\texttt{spectral}\_cube \cite{spectralcube}, 
\texttt{uvplot} \cite{uvplot}. 
\item {\bf Author contribution}: 
D.Z. reduced and analyzed the data, interpreted the results, produced the figures, and drafted the manuscript. 
S.C.C conceived, designed, and supervised the projects. 
R.G produced the tSZ model of SPT2349$-$56 for the ALMA proposal. 
R.D validated the fidelity of the tSZ decrement. 
P.A.A calculated the redshift-evolution of the Compton-$Y$ parameter and hot-gas fraction from TNG-Cluster simulations. 
S.K reduced the JWST NIRCam data. 
All authors contributed substantially to discussing the results and preparing the manuscript.
\item {\bf Author information:} The authors declare no competing interests. 
Correspondence and requests for materials should be addressed to D.Z. (\href{mailto:dzhou.astro@gmail.com}{dzhou.astro@gmail.com}).
\end{itemize}

\bibliography{sn-bibliography}

\end{document}